\documentclass{amsart}
\usepackage{amsmath}
\usepackage{amsfonts} 
\usepackage{graphics}
\usepackage{epsfig}
\usepackage{amssymb}
\usepackage{amscd}

\newtheorem{thm}{Theorem}[section]

\newtheorem{lma}[thm]{Lemma}
\theoremstyle{definition}
\newtheorem{Def}[thm]{Definition}
\newtheorem*{ack}{Acknowledgement}
\theoremstyle{remark}
\newtheorem*{rem}{Remark}

\theoremstyle{definition}
\newtheorem{ex}{Example}[section]

\numberwithin{equation}{section}
\numberwithin{figure}{section}

\begin{document}

\title[Asymptotic Expansion of 
Matrix Integral]{Lectures on the Asymptotic
 Expansion of a Hermitian Matrix Integral}
\author[Motohico Mulase]{Motohico Mulase}  
\address{
Department of Mathematics\\
University of California\\
Davis, CA 95616--8633}
\email{mulase@math.ucdavis.edu}
\date{August 18, 1997}
\subjclass{Primary: 32G15, 57R20,
81Q30. Secondary: 14H15, 
30E15, 30E20, 30F30}

\allowdisplaybreaks
\setcounter{section}{-1}
\begin{abstract}
In these lectures three different methods of 
computing the asymptotic expansion of a Hermitian
matrix integral is presented. The first one is a 
combinatorial method using Feynman diagrams. This leads
us to the generating function of the reciprocal of the 
order of the automorphism group of 
a tiling of a Riemann
surface. 
The second method is based on the classical analysis
of orthogonal polynomials. A rigorous asymptotic
method is established, and a special case of the
matrix integral is computed in terms of the 
Riemann $\zeta$-function. 
The third method 
is derived from 
a formula for the $\tau$-function solution to the 
KP equations. This method leads us to a new
class of solutions of the KP equations that are
 \emph{transcendental},
in the sense that they cannot be obtained by the 
celebrated Krichever construction and its generalizations
based on algebraic geometry of vector bundles on
Riemann surfaces. In each case a mathematically
rigorous way of dealing with asymptotic series in an
infinite number of variables is established. 
\end{abstract}
\maketitle
\tableofcontents

\newpage

\def\exp{{\text{\rm exp}}}
\def\trace{{\text{\rm trace}}}
\def\rchi{{\hbox{\raise1.5pt\hbox{$\chi$}}}}
\def\Aut{{\text{\rm Aut}}}
\def\ord{{\text{\rm ord}}}

\section{Introduction}\label{introduction}

The purpose of these lectures is to explain  three
different methods of calculation of the asymptotic
expansion of a Hermitian matrix integral. 

The first method is a combinatorial one using
the technique of Feynman diagram expansion. This method
leads us directly to the connection
between  the matrix integrals
and the moduli spaces of pointed Riemann surfaces
\cite{Harer-Zagier}, \cite{Kontsevich}, 
\cite{Penner}, \cite{Witten}.
The second method is the classical asymptotic 
analysis of orthogonal polynomials. It allows us
to compute the integral explicitly in the special case
known as the Penner Model, 
which is related to the Euler characteristic of the
moduli spaces of Riemann surfaces. 
We will see that the values are expressed in terms 
of the Riemann  zeta function. Except for this special
case,  the integral in general reduces
to a Selberg integral which is not explicitly
computable. However, through the fact that
the Hermitian matrix integral satisfies the KP
equations, we give another expression of the
asymptotic expansion as a $\tau$-function of the
KP equations.

The Hermitian matrix integral thus connects three
different worlds of mathematics: the moduli theory of
Riemann surfaces through combinatorics,  the
Riemann  zeta function through classical
asymptotic analysis, and the theory of integrable
systems through 
$\tau$-functions of the KP
equations. We explain  these relations in this
article, however, no attempt will be made to 
give any conceptual or geometric explanation 
why the KP equations are related to the topology of
moduli spaces of pointed Riemann surfaces.

Riemann's collected work is a great source
of imagination to a mathematician. The Riemann
theta functions were introduced in 
his monumental paper
\emph{Theorie der Abel'schen Functionen} that
was published in Crelle's journal in 1857. Two years later
he published a paper on the prime number distribution
where he studied the property of the 
 zeta function as a complex analytic function. 
These papers
are unrelated, but we note that his proof of the
functional equation of the zeta function is based on the
transformation property of a Jacobi theta function
with respect to the Jacobi imaginary transform
$\tau \mapsto -1/\tau$. The Jacobi theta functions
are the 1-dimensional version of the Riemann
theta functions, and 
the Jacobi imaginary transform is a special case of
more general modular transforms in the moduli
parameters.
The coincidental equivalence between the functional
equation of the Riemann zeta function and the modular
invariance of a theta function is mysterious. How much
more did Riemann know about the relations between 
these two types of functions?

In the following sections
we explore another relation between these two types of
functions. 
 The way we will encounter the moduli spaces of 
Riemann surfaces  is quite different
from Riemann's in the above mentioned
paper of 1857. They appear very naturally in the 
asymptotic expansion of  Hermitian matrix integrals, 
which can be considered as  a kind of generalization
of the Riemann theta functions. We know that 
Riemann theta functions associated with 
Riemann surfaces are characterized as  
finite-dimensional solutions to the
system of KP equations \cite{AD},
 \cite{Mulase1984}, \cite{Mumford1978},
\cite{Shiota}. The matrix integrals that
we will investigate in this article satisfy again the same KP
equations, though this time they are truly 
infinite-dimensional solutions
\cite{Mulase1994}.

Using
a combinatorial and number-theoretic 
method,  Harer and Zagier
\cite{Harer-Zagier} obtained a formula for the Euler
number of the moduli space of pointed Riemann
surfaces
(defined as an algebraic stack or an orbifold)
 in terms of the Riemann  zeta function.
Later an analytic method of calculating
the asymptotic expansion of a special Hermitian matrix
integral was proposed by Penner \cite{Penner}. 
He discovered
that the coefficients of the asymptotic series are
given in terms of special values of 
the Riemann  zeta function. 
Penner's proposed computation coincides with the 
formula of Harer and Zagier,
except for the subtle point of
giving an ordering to the set of marked points or not. 
The calculation of the asymptotic expansion of 
the Penner model has been rigorously
performed \cite{Mulase1995}.
The theorem of
 Harer and  Zagier gives an amazing relation
between the Riemann zeta function and the Riemann
theta functions, if we think the latter to be essentially
related to the moduli spaces of Riemann surfaces.

We add to this link yet another player: the KP equations. 
The observation  
\cite{Mulase1994b} that the Hermitian
matrix integral is a continuum soliton solution to the 
KP equation is suggestive from the geometric point
of view. Soliton solutions represent singular Riemann
surfaces with rational double points. When we increase the
number of singularities to continuum infinity,   the
$\tau$-function of the soliton solution converges to
a Hermitian matrix integral that has the information 
of the Euler characteristic of the moduli spaces of 
pointed Riemann surfaces.  We do not know why. 

Many explicit formulas for solutions of the KP equations
have been established. All these solutions are based
on the one-to-one correspondence between 
certain class of solutions of the KP equations and
a set of geometric data consisting of an 
 arbitrary irreducible
algebraic curve, which can be singular as well, and
a torsion-free sheaf defined on it 
\cite{Mulase1990}. Let us call a
solution to the KP equations \emph{transcendental}
if it does not correspond to any algebraic curve. 
How can we construct a transcendental solution, then?
An answer has been obtained by an accident. 
It turns out that 
the Hermitian matrix integrals we deal with
in this article are  transcendental 
solutions of the KP equations. This is closely
related to the unexpected $sl(2)$ stability condition of the
points of the infinite-dimensional Grassmannian of
Sato \cite{Sato} that correspond to the matrix
integrals. Again we do not have any satisfactory
explanation why the KP equations, the $sl(2)$ stability 
condition, and the Euler characteristic of the moduli
spaces of pointed Riemann surfaces are
related.  The last section 
 is devoted to this topic.

The  organization of the article is  as follows. In
Section~\ref{feynman} we explain the technique of
the Feynman diagram expansion through a 
toy model. A Feynman diagram is a 
kind of \emph{graph}, but
the notion of the automorphism group of a Feynman 
diagram is different from the usual graph theoretic
automorphism. This topic is carefully treated in this
section. Section~\ref{matrix} is devoted to
explaining the ribbon graph expansion of a Hermitian
matrix integral. The mathematical method of 
dealing with  asymptotic series in
an infinite number of variables is also explained in this
section. The Penner model is rigorously calculated in
Section~\ref{asymptotic}, following the idea of
\cite{Mulase1995}. The value we obtain is the 
Euler characteristic of the moduli spaces of pointed 
Riemann surfaces calculated by Harer and Zagier, but
we will not go into the moduli theory in this article. 
The third expression of the asymptotic expansion
of the Hermitian matrix integral  is computed by
using the formula for the $\tau$-function solution
to the KP equations in Section~\ref{kp}. This
solution is transcendental, which is proved 
in Section~\ref{grass}
from the $sl(2)$ stability condition of the point
of the Grassmannian that corresponds to the
Hermitian matrix integral.  The last two
sections contain our new results, including
Theorem~\ref{thm4.2} and
Theorem~\ref{thm: transcendental solution},
which were presented in the UIC Workshop
in 1997.

\begin{ack}
This article is based 
on the series of  lectures delivered by the  author as 
 graduate courses at 
Kyoto University (1994, 1995),
Mathematical Society of Japan Summer Institute
for Youth (1995), Humboldt Universit\"at zu Berlin
(1995, 1996), 
and the University of California, Davis (1994, 1996).
 He thanks the
organizers and the enthusiastic audience of
these courses, in particular,  
Mikio and Yasuko Sato, Takahiro Shiota, and
Kenji Ueno of Kyoto,
Thomas Friedrich,
Herbert Kurke, and Ines Quandt of 
Berlin,
 and Michael Pencava and Craig Tracy of Davis, for
encouragements and  valuable comments. 
The author's  special
thanks are due to Laura Loos who went through the
earlier version of the lecture notes  and
made useful comments and suggestions
that are incorporated in this article.
\end{ack}

\section{Feynman diagram expansion of a toy model}
\label{feynman}
Let us start with a simple integral:
\begin{equation}\label{eq1.1}
\int_{-\infty} ^{\infty} e^{-x^2/2} dx = \sqrt{2\pi} .
\end{equation}
According to Lord Kelvin, a mathematician is one to whom
that is as obvious as that twice two makes four is to you.
However, the usual proof of this formula using polar coordinates
of a plane
 is \emph{really trivial}, and it is hardly a good
qualification for a mathematician.
It is plausible that Lord Kelvin had in mind  a proof  using
functions only in one variable and appealing to
an infinite product expansion of trigonometric functions,
that requires reasonably deep knowledge of function
theory.

The integral we consider  is a variation of
(\ref{eq1.1}):
\begin{equation}\label{eq1.2}
Z(t) = \int_{-\infty} ^{\infty} e^{-x^2/2 } e^{t\cdot x^4/4!}
\frac{dx}{\sqrt{2\pi}}  .
\end{equation}
We want to know the integral $Z(t)$ as a function of $t$.
 Since
$$
\big| e^{t\cdot x^4/4!} \big| = e^{Re(t)\cdot x^4 /4!}
$$
for every $x\in {\mathbb{R}}$, the integral converges to make
$Z(t)$  a holomorphic function in $t$ for $Re(t) < 0$. 
Unfortunately there is no analytic
method to give a simple closed formula like (\ref{eq1.1})
for (\ref{eq1.2}), so we need a different approach. 
Since a holomorphic function
defined on a domain is completely determined by its
convergent Taylor expansion at a point in the domain, we 
can try to find a convergent power series expansion of
$Z(t)$. But here again we encounter the same
problem, and the only  thing we can do is restricted
to the power series expansion of $Z(t)$ at $t=0$. 
 At a boundary point of the domain where
the function is not holomorphic,
there is no longer a
Taylor expansion, but we still have a useful
power series expansion called an
\emph{asymptotic expansion}.

\begin{Def}Let $\Omega$ be an open
domain of the complex plane ${\mathbb{C}}$ having the
origin $0$ on its boundary, and let $h(z)$ be a
holomorphic function defined on $\Omega$. A formal
power series
$$
\sum_{v=0} ^{\infty} a_{v} z^{v}
$$
is said to be an \emph{asymptotic} expansion of $h(z)$
on $\Omega$ at $z=0$ if
\begin{equation}\label{eq1.3}
\lim_{\substack{z\rightarrow 0 \\ z\in \Omega}}
\frac{h(z) - \sum_{v = 0} ^m a_{v} z^{v}}
{z^{m+1}} = a_{m+1}
\end{equation}
holds for all $m\ge 0$.
\end{Def}

\noindent
If $h(z)$ happens to be holomorphic at $z=0$,
then the Taylor series expansion of $h(z)$ at the origin
is by definition an asymptotic expansion.
Formula (\ref{eq1.3}) shows that
if $h(z)$ admits an asymptotic expansion, then
it is unique. However, we cannot recover the
original holomorphic function from its asymptotic
expansion. Let us compute the asymptotic
expansion of $e^{1/z}$ defined
on a domain  
\begin{equation}\label{omegaepsilon}
\Omega_{\epsilon} = 
\{z\in{\mathbb{C}} | \pi/2 + \epsilon < \arg(z)
<3\pi/2 - \epsilon\}
\end{equation}
 for a small $\epsilon >0$.  Since
$$
\lim_{\substack{ z\rightarrow 0 \\ z\in \Omega_\epsilon}}
\frac{e^{1/z} - 0}
{z^{m+1}} = 0
$$
for any $m\ge 0$, the asymptotic expansion
of $e^{1/z}$ at the origin is the $0$-series. Thus the
asymptotic expansion does not recognize the difference
between $e^{1/z}$ and the $0$-function. We will use this
fact many times in Section~\ref{asymptotic} 
when we compute the
Penner model. This example also shows us that even
when $h(z)$ is not holomorphic at $z=0$, its
asymptotic expansion can be a convergent power series.

\begin{figure}[hbt]
\centerline{\epsfig{file=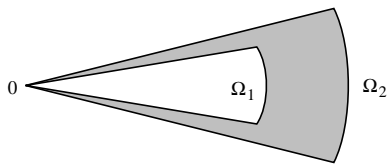, width=2in}}
\caption{Domains $\Omega_1\subset \Omega_2$}\label{figasymptotic} 
\end{figure}

To indicate that the asymptotic expansion of a holomorphic
function is  \emph{not equal} to the original function,
we use the following notation:
$$
{{\mathcal{A}}}\big(h(z)\big) =
\sum_{v=0} ^{\infty} a_{v} z^{v} .
$$
If two holomorphic functions $h(z)$ and $f(z)$
defined on $\Omega$ have the same asymptotic
expansion at $z= 0$, then we write
$$
h(z) \overset{A}\equiv f(z) .
$$
Thus $0 \overset{A}\equiv e^{1/z}$ at $z = 0$ as
holomorphic functions defined on the domain
$\Omega_{\epsilon}$.
For two holomorphic functions $f(z)$ and $g(z)$ defined
on $\Omega$ admitting the asymptotic expansions at 
$0$, we have
\begin{equation*}
\begin{split}
\mathcal{A}\big(f(z) + g(z)\big) 
&= \mathcal{A}\big(f(z) \big) + 
\mathcal{A}\big(g(z)\big)\\
\mathcal{A}\big(f(z)\cdot g(z)\big) 
&= \mathcal{A}\big(f(z) \big) \cdot 
\mathcal{A}\big(g(z)\big) .
\end{split}
\end{equation*}
We note that the asymptotic expansion of a holomorphic
function \emph{does} depend on the choice of the domain $\Omega$.
For example, $e^{1/z}$ does not admit any asymptotic
expansion at $z=0$ as a holomorphic function on the right half
plane. However,
if  
$$
\Omega_1 \subset \Omega_2,\qquad 0\in
\partial\Omega_1\cap\partial\Omega_2,
$$
as in Figure~\ref{figasymptotic},  and $h(z)$ has an
asymptotic expansion on $\Omega_2$ at $z=0$,
then it also admits an asymptotic expansion on $\Omega_1$
at $z=0$, which is actually the same series.

We can also define the asymptotic expansion  of a real analytic
function: if $K$ is an open interval of the real axis with $0$
as its one of the boundary points
and $h(z)$  a real analytic function on $K$,
then the same formula (\ref{eq1.3}), replacing $\Omega$ by
$K$, defines the asymptotic
expansion of $h(z)$ at $z=0$.

Now let us compute
the asymptotic expansion of
 $Z(t)$ of (1.2) as  a holomorphic function defined on 
$\Omega_{\epsilon} = 
\{t\in{\mathbb{C}} | \pi/2 + \epsilon < \arg(t)
<3\pi/2 - \epsilon\}$.  The Taylor
expansion of the exponential gives
$$
 \int_{-\infty} ^{\infty} e^{-x^2/2 } e^{t\cdot x^4/4!}
\frac{dx}{\sqrt{2\pi}} =
 \int_{-\infty} ^{\infty} e^{-x^2/2 } \sum_{v=0} ^{\infty}
\frac{1}{(4!)^v\cdot v!}\cdot  x^{4v}\cdot t^v\cdot
\frac{dx}{\sqrt{2\pi}} .
$$
The infinite
integral and the infinite sum 
we have here are not  interchangeable.
But  let's just
interchange them and see what happens:
\begin{equation}\label{eq1.4}
 \sum_{v=0} ^{\infty}
\frac{1}{(4!)^v\cdot v!} \left(
\int_{-\infty} ^{\infty} e^{-x^2/2 } \cdot  x^{4v}
\cdot \frac{dx}{\sqrt{2\pi}} \right) t^v .
\end{equation}
Note that this is a well-defined formal power series in
$t$ because the integral
$$
\int_{-\infty} ^{\infty} e^{-x^2/2 } \cdot  x^{4v}
\cdot
\frac{dx}{\sqrt{2\pi}}
$$
converges.
\begin{lma}\label{lma1.2}
The formal power series \eqref{eq1.4}
gives the asymptotic expansion
of $Z(t)$:
\begin{multline*}
{\mathcal{A}}\left( \int_{-\infty} ^{\infty} e^{-x^2/2 } \sum_{v=0} ^{\infty}
\frac{1}{(4!)^v\cdot v!}\cdot  x^{4v}\cdot t^v\cdot
\frac{dx}{\sqrt{2\pi}}
\right) \\
= \sum_{v=0} ^{\infty}
\frac{t^v}{(4!)^v\cdot v!} \left(
\int_{-\infty} ^{\infty} e^{-x^2/2 } \cdot  x^{4v}
\cdot \frac{dx}{\sqrt{2\pi}} \right)  .
\end{multline*}
\end{lma}

\noindent
Although we cannot get an \emph{equality} by
interchanging the integral and the sum
because the power series expansion
of the integrand  of Lemma \ref{lma1.2} is not
uniformly convergent
on the infinite interval $(-\infty,\infty)$,
 at least we obtain
a formula which is correct in one direction.
\begin{proof}
Using the linearity of the integral, we have
\begin{align*}
 &\int_{-\infty} ^{\infty} e^{-x^2/2 } \sum_{v=0} ^{\infty}
\frac{1}{(4!)^v\cdot v!}\cdot  x^{4v}\cdot t^v\cdot
\frac{dx}{\sqrt{2\pi}}
-
 \sum_{v=0} ^m
\frac{t^v}{(4!)^v\cdot v!}
\int_{-\infty} ^{\infty} e^{-x^2/2 } \cdot  x^{4v}
\cdot \frac{dx}{\sqrt{2\pi}} \\
= &\int_{-\infty} ^{\infty} e^{-x^2/2 } \sum_{v=m+1} ^{\infty}
\frac{1}{(4!)^v\cdot v!}\cdot  x^{4v}\cdot t^v\cdot
\frac{dx}{\sqrt{2\pi}}\\
=&t^{m+1}\int_{-\infty} ^{\infty} e^{-x^2/2 } \sum_{a=0} ^{\infty}
\frac{1}{(4!)^{m+1+a}\cdot
{(m+1+a)!}}\cdot  x^{4(m+1+a)}\cdot t^{a}\cdot
\frac{dx}{\sqrt{2\pi}} .
\end{align*}
As long as $t$ stays in 
$\Omega_{\epsilon} = 
\{t\in{\mathbb{C}} | \pi/2 + \epsilon < \arg(t)
<3\pi/2 - \epsilon\}$, we can divide the
above expression by $t^{m+1}$ and take the limit $t\rightarrow 0$,
because the integral converges. The result is the $(m+1)$-th
coefficient of the asymptotic expansion, which
proves the claim.
\end{proof}

\begin{figure}[htb]
\centerline{\epsfig{file=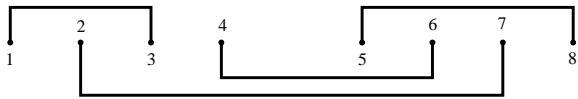, width=3in}}
\caption{Pairing Scheme}\label{fig1.2} 
\end{figure}

How can we calculate the coefficient of the expansion? The
standard technique is the following:
\begin{align*}
\int_{-\infty} ^{\infty} e^{-x^2/2 } \cdot  x^{4v}
\cdot \frac{dx}{\sqrt{2\pi}}
&=
\int_{-\infty} ^{\infty} e^{-x^2/2 } \cdot
\left. \left(\frac{d}{dy}\right)^{4v}e^{xy}\right|_{y=0}
\cdot \frac{dx}{\sqrt{2\pi}}\\
&= \left(\frac{d}{dy}\right)^{4v}
\int_{-\infty} ^{\infty} e^{-x^2/2 } \cdot
e^{xy}
\cdot \left.\frac{dx}{\sqrt{2\pi}}\right|_{y=0}\\
&=\left(\frac{d}{dy}\right)^{4v} \left.
\int_{-\infty} ^{\infty} e^{-(x-y)^2/2 } \cdot
e^{y^2/2}
\cdot \frac{dx}{\sqrt{2\pi}}\right|_{y=0}\\
&=\left.\left(\frac{d}{dy}\right)^{4v}e^{y^2/2}\right|_{y=0} ,
\end{align*}
where we have used the translational invariance of the
integral (\ref{eq1.1}).
Note that the integration is reduced to a differentiation.
All we need now is a Taylor coefficient of the exponential
function $e^{y^2/2}$, from which we obtain
\begin{equation}\label{eq1.5}
\int_{-\infty} ^{\infty} e^{-x^2/2 } \cdot  x^{4v}
\cdot \frac{dx}{\sqrt{2\pi}}  =
\left.\left(\frac{d}{dy}\right)^{4v}e^{y^2/2}\right|_{y=0}
= \frac{(4v)!}{(2v)!\cdot 2^{2v}}
=(4v -1)!! ,
\end{equation}
where the double factorial is defined by
$$
(2n-1)!! = (2n-1)\cdot (2n-3)\cdot (2n-5)
\cdots 5\cdot 3\cdot 1 .
$$
The quantity (\ref{eq1.5}) has a combinatorial meaning. Let us denote
the differential operator $d/dy$ by a dot $\bullet$. We have
$4v$ dots attacking the fort $e^{y^2/2}$. Since $y$ is set equal to 
$0$ after the operation, if only one dot attacks the fort, the result
would be just $0$:
$$
\left.\frac{d}{dy} \ e^{y^2/2}\right|_{y=0} = y\  e^{y^2/2}\big|_{y=0} = 0 .
$$
To obtain a nonzero result, the dots have to attack the fort
by pairs:
$$
\left.\left(\frac{d}{dy}\right)^2
 e^{y^2/2}\right|_{y=0} = y^2\  e^{y^2/2}\big|_{y=0}
+e^{y^2/2}\big|_{y=0}= 1 .
$$
Noting that the result we get by the paired attack is $1$,
we conclude that the value of the integral (or the
differentiation) (\ref{eq1.5}) is
equal to
\begin{multline*}
{\text{\emph{The number of ways of making  $2v$ pairs
out of $4v$ dots}}}\\
\begin{aligned}
=&\binom{4v}{2}\binom{4v-2}{2}\binom{4v-4}{2}
\cdots \left.\binom{4}{2}\binom{2}{2}\right/(2v) !\\
=&\frac{4v(4v-1)}{2}\cdot\frac{(4v-2)(4v-3)}{2}
\cdots \frac{4\cdot 3}{2}\cdot\left.\frac{2\cdot 1}{2}\right/(2v)!\\
=& \frac{(4v)!}{(2v)!\cdot 2^{2v}} .
\end{aligned}
\end{multline*}
These pairs can be visualized by a diagram like 
Figure~\ref{fig1.2}. Let us call such a diagram a
\emph{pairing scheme}. Thus (\ref{eq1.5}) gives the number 
of pairing schemes of $4v$ dots.
An example of a pairing scheme of $8=4\times 2$ dots is given 
in Figure~\ref{fig1.2}. 

\begin{figure}[htb]
\centerline{\epsfig{file=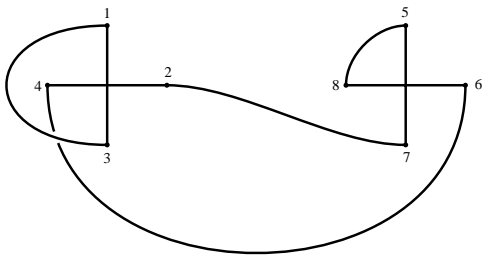, width=2.5in}}
\caption{Feynman Diagram}\label{fig1.3} 
\end{figure}

The coefficient of the asymptotic expansion of Lemma
\ref{lma1.2} has an
extra factor of $1/(4!)^{v}\cdot {v}!$. How can we interpret
it combinatorially? Here enters the idea of Feynman diagrams.
The $4v$ dots are grouped into $v$ sets of $4$ dots.
Let us replace each set of $4$ dots by a
\emph{cross}, identifying the four dots with
the four endpoints of the cross. Then the pairing scheme
changes into a \emph{Feynman diagram}, 
as shown in Figure~\ref{fig1.3},
by connecting the endpoints according to the pairing
rules.
This is an example of a \emph{graph}. We use this word
for a CW complex like Figure~\ref{fig1.3} in this article.
A graph $\Gamma = (V, E, i)$
consists of a finite set $V$ of \emph{vertices}, 
a finite set 
$E$ of \emph{edges}, and the incidence relation $i$ of
vertices and edges. The number of half-edges
coming out of a vertex 
of a graph  is called
the \emph{degree}  of the
vertex. A \emph{degree} $d$ graph
is a graph whose vertices have the same degree
$d$. A
degree $3$ graph is also called a \emph{trivalent}
graph. 
 The \emph{order} 
of the graph $\Gamma$ is the number of vertices $|V|$ 
of $\Gamma$.

When we make the
Feynman diagram $\Gamma$ from a pairing scheme, we 
consider the center of a cross as a vertex 
 and a pairing of dots as an edge of the graph $\Gamma$.
Figure~\ref{fig1.3} is thus 
considered as a degree $4$ graph of order
$2$. 

As a  graph we can interchange the $v$ crosses
freely, and in each cross we can place  the four
edges
in any way we want, as long as
\emph{the strings are attached}.
The degrees of freedom for these moves are exactly
$(4!)^{v}\cdot v!$. Thus we (tentatively) conclude that
\emph{the $v$-th coefficient of the asymptotic expansion
of the integral $Z(t)$ is the number of degree $4$
graphs of order $v$}. As an example, let us
compute the simplest case $v = 1$. From the
above considerations, the number of degree
$4$ graphs with one vertex should be
$$
\frac{(4-1)!!}{4!} = \frac{1}{8} .
$$
But this is impossible! What went wrong?

\begin{figure}[htb]
\centerline{\epsfig{file=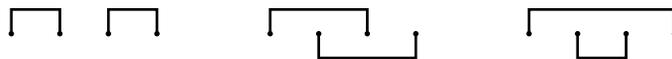, width=3.5in}}
\caption{Pairing Schemes of 4 Dots}\label{fig1.4} 
\end{figure}

The number of different pairing schemes of
$4$ dots is three,
as  in Figure~\ref{fig1.4}.
When we factored out (\ref{eq1.5}) by $(4!)^{v}\cdot v!$, we assumed
that interchanging the $v$ crosses and renumbering
each edge of a cross would lead to a different pairing
scheme that still corresponds to the same graph.
In other words, we assumed that the group 
${\mathfrak{S}}_{v}
\rtimes ({\mathfrak{S}}_4)^{v}$
acts on the set of all pairing schemes freely,
where ${\mathfrak{S}}_n$ denotes the
permutation group of $n$ letters, and the product is the
semi-direct product of two factors
with $({\mathfrak{S}}_4)^{v}$ as its normal
subgroup. But as we see clearly from the above example,
some pairing schemes are stable under
the action of non-trivial permutations. The isotropy
group that stabilizes
 any of the three pairing schemes of Figure~\ref{fig1.4}
is $({\mathbb{Z}}\big/ 2{\mathbb{Z}})^3$.

Since our graphs are constructed from pairing schemes,
we  define the \emph{automorphism group}
 in the following manner:

\begin{Def} The automorphism group
$\Aut(\Gamma)$ of a graph $\Gamma$ is
the isotropy subgroup $I_P$ of ${\mathfrak{S}}_{v}
{\rtimes}({\mathfrak{S}}_4)^{v}$ that 
preserves the original pairing
scheme $P$. If pairing schemes $P$ and $P'$ correspond
to $\Gamma$, then the  isotropy groups $I_P$
and $I_{P'}$ are conjugate to one another in
${\mathfrak{S}}_{v}
\rtimes ({\mathfrak{S}}_4)^{v}$. 
Therefore, as an abstract group,
$\Aut(\Gamma)$ is well-defined.
\end{Def}

\begin{rem} 
 Our definition of $\Aut(\Gamma)$
\emph{does not} coincide with the traditional 
graph theoretic definition of
automorphism.
\end{rem}

The correct interpretation of $1/8$ is then
$1/|\Aut(\Gamma)|$, where $\Gamma$ in this case
is the degree $4$
graph with only one vertex,
and we denote by $|\Aut(\Gamma)|$ the order of
the group. More generally, we can interpret
the formal power series (\ref{eq1.4}) as 
a summation over the set
of all pairing schemes modulo the group ${\mathfrak{S}}_{v}
\rtimes ({\mathfrak{S}}_4)^{v}$, which is 
equivalent to the set of all degree
$4$ graphs. The contribution of a graph $\Gamma$
is modified
by the weight of $1/|\Aut(\Gamma)|$.
Summarizing, we have established:
\emph{The asymptotic expansion of the integral 
$Z(t)$ is given by}
$$
{{\mathcal{A}}}\left(  \int_{-\infty} ^{\infty} e^{-x^2/2 } e^{t\cdot x^4/4!}
\frac{dx}{\sqrt{2\pi}} \right)
=\sum_{v=0} ^{\infty}
\left(
\sum_{\substack{{\text{degree $4$ graph}} \\
\Gamma{\text{ of order $v$}}}}
\frac{1}
{|\Aut(\Gamma)|}
\right)\cdot t^{v}  .
$$
Since the number of degree 
$4$ graphs with a fixed number
of vertices is finite, the right hand side of the above formula is
a well-defined element of the power series ring
${\mathbb{Q}}[[t]]$.
The degree of each vertex of  the graph is $4$,
which is due to the power $4$ in the
exponent of the integral. The same argument thus establishes
\begin{thm}\label{thm1.4}
The  asymptotic formula 
$$
{{\mathcal{A}}}\left(  \int_{-\infty} ^{\infty} e^{-x^2/2 } 
e^{t\cdot x^{2j}/(2j)!}
\frac{dx}{\sqrt{2\pi}} \right)
=\sum_{v=0} ^{\infty}
\left(
\sum_{\substack{{\text{degree $2j$ graph }} \\
\Gamma{\text{ of order  $v$}}}}
\frac{1}
{|\Aut(\Gamma)|}
\right) \cdot t^{v} \in {\mathbb{Q}}[[t]]
$$
holds for an arbitrary
$j \ge 2$.
\end{thm}

\noindent
We can consider more general graphs with the integral
\begin{equation}\label{eq1.6}
Z(t_1,t_2,\cdots,t_{2m})=\int_{-\infty} ^{\infty} e^{-x^2/2 }
{\exp}\left(\sum_{j=1} ^{2m} \frac{t_j}{j!}  x^j\right)
\frac{dx}{\sqrt{2\pi}}  ,
\end{equation}
where $m\ge 2$ is an integer. The integral converges
if $t_{2m}$ is in the domain $\Omega_\epsilon$
of (\ref{omegaepsilon}) and determines a
holomorphic function on
$$
(t_1,t_2,\cdots,t_{2m})
\in {\mathbb{C}}^{2m-1}
\times \Omega_\epsilon .
$$
We can expand $Z(t_1,t_2,\cdots,t_{2m})$ as a Taylor
series in $(t_1,t_2,\cdots,t_{2m-1})
\in {\mathbb{C}}^{2m-1}$ and as an
asymptotic series in $t_{2m}\in \Omega_\epsilon$
 at the origin. Fix a
value of $t_{2m}\in \Omega_\epsilon$. Then
$$
{\exp}\left({t_{2m}}\cdot x^{2m}/{(2m)!}\right)
$$
acts as a uniformizing factor so that the
power series expansion  of the
integrand in terms of $x$ converges uniformly on
$(-\infty,\infty)$ for all values of $t_1$, $t_2$,
$\cdots$, $t_{2m-1}\in{\mathbb{C}}$.  Therefore, we can
interchange the infinite integral and the infinite sums:
\begin{align*}
& Z(t_1,t_2,\cdots,t_{2m})=\int_{-\infty} ^{\infty} e^{-x^2/2 }
{\exp}\left(\sum_{j=1} ^{2m} \frac{t_j}{j!}  x^j\right)
\frac{dx}{\sqrt{2\pi}}\\
&=\int_{-\infty} ^{\infty} e^{-x^2/2 }
{\exp}\left(\frac{t_1}{1!}  x\right)
\cdots
{\exp}\left(\frac{t_{2m-1}}{(2m-1)!}  x^{2m-1}\right)
\cdot
{\exp}\left(\frac{t_{2m}}{(2m)!}  x^{2m}\right)
\frac{dx}{\sqrt{2\pi}}\\
&=\int_{-\infty} ^{\infty} e^{-x^2/2 }
\left(\sum_{v_1=0} ^\infty
\frac{{t_1}^{v_1}}{v_1!\cdot (1!)^{v_1}}\right)
\cdots
\left(\sum_{v_{2m-1}=0} ^\infty
\frac{{t_{2m-1}}^{v_{2m-1}}}
{v_{2m-1}!\cdot ((2m-1)!)^{v_{2m-1}}}\right)\\
&\quad \times
\left(\sum_{v_{2m}=0} ^\infty
\frac{{t_{2m}}^{v_{2m}}}
{v_{2m}!\cdot ((2m)!)^{v_{2m}}}\right)
x^{v_1 + 2v_2 + \cdots +(2m)v_{2m}}
\frac{dx}{\sqrt{2\pi}}\\
&=\left(\sum_{v_1=0} ^\infty
\frac{{t_1}^{v_1}}{v_1!\cdot (1!)^{v_1}}\right)
\cdots
\left(\sum_{v_{2m-1}=0} ^\infty
\frac{{t_{2m-1}}^{v_{2m-1}}}
{v_{2m-1}!\cdot ((2m-1)!)^{v_{2m-1}}}\right)\\
&\quad\times\int_{-\infty} ^{\infty} e^{-x^2/2 }
\left(\sum_{v_{2m}=0} ^\infty
\frac{{t_{2m}}^{v_{2m}}}
{v_{2m}!\cdot ((2m)!)^{v_{2m}}}\right)
x^{v_1 + 2v_2 + \cdots +(2m)v_{2m}}
\frac{dx}{\sqrt{2\pi}} .
\end{align*}
We already know that
\begin{multline*}
{\mathcal{A}}\left(
\int_{-\infty} ^{\infty} e^{-x^2/2 }
\left(\sum_{v_{2m}=0} ^\infty
\frac{{t_{2m}}^{v_{2m}}}
{v_{2m}!\cdot ((2m)!)^{v_{2m}}}\right)
x^{v_1 + 2v_2 + \cdots +(2m)v_{2m}}
\frac{dx}{\sqrt{2\pi}}\right)\\
=\sum_{v_{2m}=0} ^\infty
\frac{{t_{2m}}^{v_{2m}}}
{v_{2m}!\cdot ((2m)!)^{v_{2m}}}
\int_{-\infty} ^{\infty} e^{-x^2/2 }
x^{v_1 + 2v_2 + \cdots +(2m)v_{2m}}
\frac{dx}{\sqrt{2\pi}}
\end{multline*}
at $t_{2m}=0$ when the top integral is considered to be a
holomorphic function in $t_{2m}\in \Omega_\epsilon$ .
Therefore, we have
\begin{multline}\label{eq1.7}
 {{\mathcal{A}}}\left(\int_{-\infty} ^{\infty} e^{-x^2/2 }
{\exp}\left(\sum_{j=1} ^{2m} \frac{t_j}{j!}  x^j\right)
\frac{dx}{\sqrt{2\pi}}\right)\\
=\sum_{v_1=0} ^\infty
\frac{{t_1}^{v_1}}{v_1!\cdot (1!)^{v_1}}
\cdots
\sum_{v_{2m}=0} ^\infty
\frac{{t_{2m}}^{v_{2m}}}
{v_{2m}!\cdot ((2m)!)^{v_{2m}}}
\int_{-\infty} ^{\infty} e^{-x^2/2 }
x^{v_1 + 2v_2 + \cdots +(2m)v_{2m}}
\frac{dx}{\sqrt{2\pi}} .
\end{multline}
We now apply the Feynman diagram expansion to
the above integral.  First, we have
$$
\int_{-\infty} ^{\infty} e^{-x^2/2 }
x^{v_1 + 2v_2 + \cdots +(2m)v_{2m}}
\frac{dx}{\sqrt{2\pi}}
=\left.
 \left(\frac{d}{dy}\right)^{v_1 + 2v_2 + \cdots +(2m)v_{2m}}
e^{y^2/2}\right|_{y=0} .
$$
The pairing scheme of the dot diagram has $v_1$
sets of single dot, $v_2$ sets of double dots, $\cdots$,
and $v_{2m}$ sets of $2m$ dots. Passing to a Feynman
diagram, we have a graph with $v_j$ 
 vertices of degree $j$ for $j=1, 2, \cdots, 2m$. Thus
\begin{thm}\label{thm1.5}
We have the following asymptotic formula:
$$
{{\mathcal{A}}}\left(  \int_{-\infty} ^{\infty} e^{-x^2/2 }
\exp\left(\sum_{j=1} ^{2m} \frac{t_j}{j!}  x^j\right)
\frac{dx}{\sqrt{2\pi}} \right)
=\sum_{\substack{{\text{Graph }}\Gamma 
{\text{ with}}\\
 {\text{vertices of degree }}\le 2m}}
\frac{1}
{|\Aut(\Gamma)|}\cdot \prod_{j = 1} ^{2m}
t_j ^{v_j(\Gamma)} ,
$$
where $v_{j}(\Gamma)$ denotes the number of 
vertices of degree $j$ in $\Gamma$.
\end{thm}

\noindent
Note that the asymptotic series is a well-defined element of the
formal power series ring
$$
{\mathbb{Q}}[[t_1,t_2,\cdots,t_{2m}]] ,
$$
because there are only finitely many graphs for given
numbers $v_1(\Gamma)$, 
$v_2(\Gamma)$, $\cdots$, $v_{2m}(\Gamma)$.

\begin{Def}
The automorphism group of $\Gamma$
 is defined as the isotropy
subgroup of
$$
\prod_{j=1} ^{2m}{\mathfrak{S}}_{v_j}
\rtimes {\mathfrak{S}}_j ^{v_j}
$$
that stabilizes the pairing scheme corresponding to
 $\Gamma$.
\end{Def}

As before, the definition of $\Aut(\Gamma)$ as an
abstract group does not depend on
the particular choice of the pairing scheme corresponding to
the graph.

\begin{figure}[htb]
\centerline{\epsfig{file=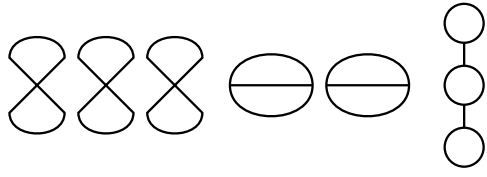, width=2.5in}}
\caption{A Disconnected Graph}\label{fig1.5} 
\end{figure}

Let us now consider the relation between 
general graphs and
connected graphs. Let $a_v$ be the number 
of arbitrary degree $j$
graphs of order $v$ and $c_v$ the number of 
\emph{connected} degree
$j$ graphs of order $v$, where $j\ge 1$ is a fixed
number. From Figure~\ref{fig1.5},
it is obvious that
\begin{equation}\label{eq1.8}
a_v = \sum_{n_1 +2n_2 +3n_3+ \cdots=v}
\frac{c_1 ^{n_1}\cdot c_2 ^{n_2}\cdot c_3 ^{n_3}\cdots}
{n_1 !\cdot n_2 !\cdot n_3 !\cdots} ,
\end{equation}
where $n_i$ is the number  of connected components with
$i$ vertices in a given graph. Formula (\ref{eq1.8}) is equivalent to
a simple functional relation in terms of generating functions:
\begin{equation}\label{eq1.9}
\sum_{v=0} ^{\infty} a_v t_j ^v = \exp\left(
\sum_{v=1} ^{\infty} c_v t_j ^v\right) ,
\end{equation}
where we use the convention that  $a_0=1$ and $c_0=0$.

In a more general case, let 
$\mathbf{v} = (v_1,v_2,\cdots,v_{2m})$
and
$$
t^{\mathbf{v}} = \prod_{j=1} ^{2m} t_j ^{v_j} .
$$
We denote by $c_{\mathbf{v}}$ the number of connected
graphs with $v_j$ vertices of degree $j$, where
$1\le j\le 2m$, and by $a_{\mathbf{v}}$ the number of all
graphs with $v_j$ vertices of degree $j$. Then we have
\begin{equation}\label{eq1.10}
\begin{split}
\sum_{\mathbf{v}} a_{\mathbf{v}} t^{\mathbf{v}}
& =\exp\left( \sum_{\mathbf{v}} 
c_{\mathbf{v}} t^{\mathbf{v}}\right)\\
& = 1 + \sum_{\mathbf{v}} c_{\mathbf{v}} t^{\mathbf{v}}
+ \frac{1}{2!} \left(\sum_{\mathbf{v}} c_{\mathbf{v}} t^{\mathbf{v}}
\right)^2 +
\frac{1}{3!}\left(\sum_{\mathbf{v}} c_{\mathbf{v}} t^{\mathbf{v}}
\right)^3
+\cdots ,
\end{split}
\end{equation}
where the $n$-th term of the right hand side counts
the number of graphs
consisting of $n$ connected components.
The factor $1/(n!)$ means that we can permute the
$n$ connected components without changing the
original graph.

The case we are
considering is slightly more complicated because the
generating function we have in Theorem~\ref{thm1.5} counts
the number of graphs with weight $1\big/|\Aut(\Gamma)|$.
The automorphism group of a graph $\Gamma$
consisting of $n$ connected components
$\Gamma_1$, $\cdots$, $\Gamma_n$
is the semi-direct product
\begin{equation}\label{eq1.11}
\Aut(\Gamma) = {\mathfrak{S}}_n 
\rtimes \prod_{j=1} ^n \Aut(\Gamma_j) .
\end{equation}
Therefore, we have
$$
\frac{1}{|\Aut(\Gamma)|}
= \frac{1}{n!}\cdot \prod_{j=1} ^n \frac{1}{|\Aut(\Gamma_j)|} .
$$
Note that the right hand side is the product of $n$ factors
following the key factor $1/(n!)$.
It shows, therefore,
that the exponential formula (\ref{eq1.10}) connecting
connected graphs and general graphs also holds in the case
we are investigating.
Thus we have  established
\begin{thm}\label{thm1.6}
The asymptotic series involving only connected graphs
is given by
\begin{multline*}
\log{{\mathcal{A}}}\left(  \int_{-\infty} ^{\infty} e^{-x^2/2 }
\exp\left(\sum_{j=1} ^{2m} \frac{t_j}{j!}  x^j\right)
\frac{dx}{\sqrt{2\pi}} \right)
\\
=\sum_{\substack{{\text{Connected graph }}\Gamma
{\text{ with}} \\
 {\text{vertices of degree }}\le 2m}}
\frac{1}
{|\Aut(\Gamma)|}\cdot \prod_{j = 1} ^{2m}
t_j ^{v_j(\Gamma)} .
\end{multline*}
\end{thm}

\noindent
We note here that the function $\log$ is \emph{not}
considered as an analytic function. It is applied to the
formal power series appearing in the right hand side of
Theorem~\ref{thm1.5} as the inverse power series of $\exp(z)$ defined
by
$$
\log(1-z) = -\sum_{n=1} ^\infty \frac{1}{n}  z^n .
$$

\section{Matrix integrals and ribbon graphs/fatgraphs}
\label{matrix}
Let ${\mathcal{H}}_n$ denote  the space of all $n\times n$
Hermitian matrices. It is an $n^2$-dimensional
Euclidean space with a metric
$$
\sqrt{\trace  (X-Y)^2} , \qquad X,Y\in {\mathcal{H}}_n .
$$
The standard volume form on ${\mathcal{H}}_n$,
which is compatible
with the above metric, is given by
$$
d\mu(X) =
dx_{11}\wedge dx_{22}\wedge\cdots
\wedge dx_{nn}\wedge
\left(\bigwedge_{i<j}d(Re  x_{ij})\wedge
d(Im  x_{ij})\right)
$$
for $X=[x_{ij}]\in {\mathcal{H}}_n$.
It is important to note that the metric
and the volume form of ${\mathcal{H}}_n$ are invariant under
the conjugation $X\longmapsto UXU^{-1}$
by a unitary matrix $U\in U(n)$.
The main subject of this section
is  the \emph{Hermitian matrix integral}
\begin{equation}\label{eq2.1}
Z_n(t,m) = \int_{{\mathcal{H}}_n} \exp\left(- 
\frac{1}{2}  \trace  (X^2)\right) \exp\left(\trace 
\sum_{j = 3} ^{2m}
\frac{t_j }{j} X^j \right)   \frac{d\mu(X)}{N} ,
\end{equation}
where  
\begin{equation}\label{eqnorm}
N = \int_{{\mathcal{H}}_n} \exp\left(- 
\frac{1}{2}  \trace  (X^2)\right)  d\mu(X)
= 2^{n/2}\cdot \pi^{n^2/2}  
\end{equation}
is a normalization constant to make $Z_n(0,m) = 1$.
We note that
$Z_n(t,m)$ is a holomorphic function
in  $(t_3, t_4, \cdots, t_{2m-1})\in
{\mathbb{C}}^{2m-3}$
and  
$$t_{2m}\in \Omega_\epsilon
=\{t\in\mathbb{C} | \pi/2+\epsilon
<\arg(t)<3\pi/2-\epsilon\}
$$ ($\epsilon >0$),
because the dominating term
$\trace  (X^{2m})$ is positive definite on ${\mathcal{H}}_n$.
Thus we can expand $Z_n(t,m)$ as a convergent power
series in $t_3$, $t_4$, $\cdots$, $t_{2m-1}$ about
$0$, and as an asymptotic series in $t_{2m}$
at $t_{2m}=0$. 

We also note here that we do not 
include the $t_1$ and $t_2$ terms in the integral
because of our interests in topology, which will
become clearer as we proceed. From the point of view
of graphs, we do not allow degree $1$ and $2$ vertices
in this section. 

\begin{figure}[htb]
\centerline{\epsfig{file=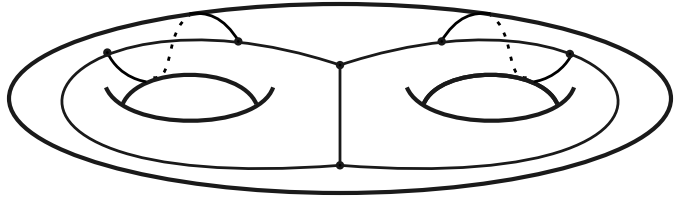, width=3.5in}}
\caption{Graph on a Surface}\label{fig2.1} 
\end{figure}

Corresponding to the fact that the integral (\ref{eq2.1}) has
richer structure than (\ref{eq1.7}), the Feynman diagrams
appearing in the asymptotic expansion of $Z_n(t,m)$
have more information than just a graph as in Theorem~\ref{thm1.5}.
As we are going to see below, the new information
we have from the Hermitian matrix integral
 is that \emph{the graph is drawn on a
compact oriented surface}. Suppose we have such a graph
drawn on an oriented surface, as in Figure~\ref{fig2.1}.

\begin{figure}[htb]
\centerline{\epsfig{file=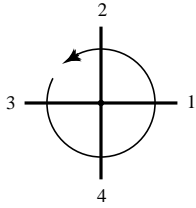, width=1in}}
\caption{Cyclic Order of Edges}\label{fig2.2} 
\end{figure}

Locally at each vertex of the graph, the orientation of the
surface gives rise to a cyclic order of the edges coming out of
the vertex, as shown in Figure~\ref{fig2.2}.

\begin{figure}[htb]
\centerline{\epsfig{file=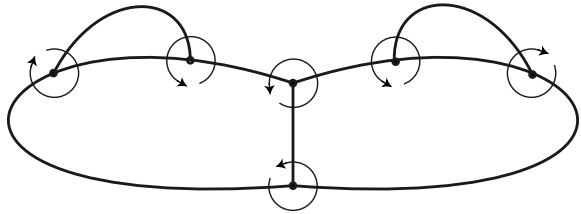, width=3in}}
\caption{Ribbon Graph}\label{fig2.3} 
\end{figure}

A graph drawn on a surface  thus gives a graph with a
cyclic order of edges at each vertex. An 
example, that is corresponding to Figure~\ref{fig2.1},
is shown in
Figure~\ref{fig2.3}.

Note that two circles are reversed
in Figure~\ref{fig2.3}, corresponding to
the fact that two edges of the graph of Figure~\ref{fig2.1}
go around the back side of the surface.

\begin{figure}[htb]
\centerline{\epsfig{file=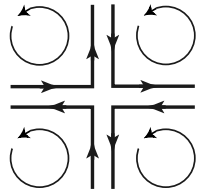, width=1in}}
\caption{Oriented Crossroad}\label{fig2.4} 
\end{figure}

Conversely, suppose we have a connected graph
$\Gamma_{rib}$ with a cyclic order of edges
assigned to each vertex. To indicate that we have the extra
information of cyclic order at each vertex, we use
$\Gamma_{rib}$ and distinguish it
from the underlying graph $\Gamma$.
We can construct a compact oriented
surface $C(\Gamma_{rib})$
canonically such that the graph $\Gamma$
is drawn on it, as follows. First, the graph around each vertex
can be drawn on a positively oriented plane that is compatible
with the cyclic order. Next we \emph{fatten} the local part
of the graph into a crossroad of multiple intersection.
The orientation of the plane defines an orientation on each
sidewalk of the crossroad, as in Figure~\ref{fig2.4}.

\begin{figure}[htb]
\centerline{\epsfig{file=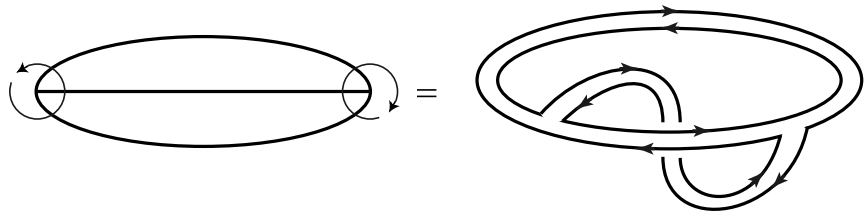, width=4.5in}}
\caption{Ribbon Graph and Surface with
Boundary}\label{fig2.5} 
\end{figure}

The roads are connected to the other parts of the graph, with
matching orientation on the sidewalks. Then we obtain an
oriented surface with boundary. Figure~\ref{fig2.5} shows
such a surface with boundary.

Let $b(\Gamma_{rib})$ denote the number of boundary components
of this oriented surface ($=$ fattened graph) made out of
$\Gamma_{rib}$. From the construction,
each boundary component has
a unique orientation compatible with that of the
fattened graph. Thus a boundary component is indeed
an oriented circle, which we also call
a \emph{boundary circuit}.
 So we can attach an oriented $2$-dimensional disk
to each boundary component of the fattened graph to construct
a compact oriented surface $C(\Gamma_{rib})$.

\begin{figure}[htb]
\centerline{\epsfig{file=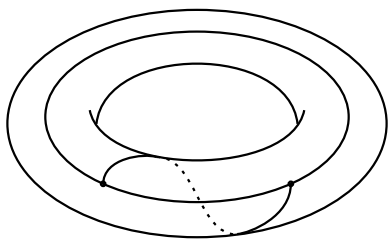, width=2in}}
\caption{3-valent Graph on a Torus}\label{fig2.6} 
\end{figure}

\begin{Def}\label{def2.1}
A ribbon graph (or a fatgraph)
is a graph with a cyclic order
of edges assigned to each vertex.
\end{Def}
\noindent
The ribbon graph
$\Gamma_{rib}$
of Figure~\ref{fig2.5} has only one boundary
component, and the resulting compact surface
$C(\Gamma_{rib})$
is the 2-torus on which the underlying graph
$\Gamma$ is drawn (Figure~\ref{fig2.6}).

We have shown that every graph drawn on an oriented surface
is a ribbon graph, and conversely,
that every connected ribbon graph
$\Gamma_{rib}$
gives rise to a canonical compact oriented surface
$C(\Gamma_{rib})$
on which the underlying graph is drawn.
 The attached boundary
disks and the underlying graph $\Gamma$
 give a cell-decomposition of
$C(\Gamma_{rib})$.

\begin{lma}\label{lma2.2}
Let $\Gamma_{rib}$ be a connected
ribbon graph with vertices of degree
$\ge 3$, and $v_j(\Gamma)$ denote the number of
 vertices of the underlying graph $\Gamma$
of degree $j$.
Then the genus $g(C(\Gamma_{rib}))$ of the
canonical oriented surface $C(\Gamma_{rib})$
associated with
$\Gamma_{rib}$ is
computed by the following formula:
$$
2-2g(C(\Gamma_{rib})) = \sum_{j\ge 3} 
v_j(\Gamma) - \frac{1}{2} 
\sum_{j\ge 3} j\cdot v_j(\Gamma) + s(\Gamma_{rib}) .
$$
\end{lma}
\begin{proof} The total number of vertices of the cell-decomposition
is given by $\sum_{j\ge 3} v_j(\Gamma)$. Since each edge is bounded
by two vertices (possibly the same), 
 the number of edges is given by
$\frac{1}{2}  \sum_{j\ge 3} j\cdot v_j(\Gamma)$. By construction,
$b(\Gamma_{rib})$ is the number of $2$-cells. Thus the
 Euler characteristic of a compact surface gives the above
formula.
\end{proof}

\begin{figure}[htb]
\begin{multline*}
\raisebox{-3mm}{\epsfig{file=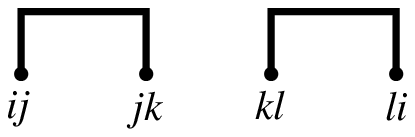, width=1.2in}} = 
\delta_{ik}\delta_{jj}\delta_{ki}\delta_{\ell \ell}\\
\raisebox{-4mm}{\epsfig{file=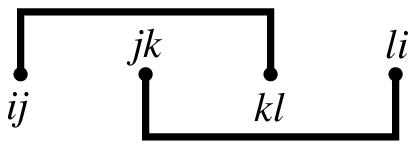, width=1.2in}} = 
\delta_{i\ell}\delta_{jk}\delta_{ji}\delta_{k\ell}\\
\raisebox{-4mm}{\epsfig{file=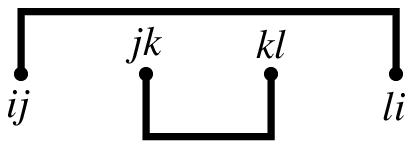, width=1.2in}} = 
\delta_{ii}\delta_{j\ell}\delta_{j\ell}\delta_{kk}
\end{multline*}
\caption{Pairing Contribution}\label{fig2.85}
\end{figure}

To see how ribbon graphs appear in the
matrix integral, let us consider a
simple example:
$$
\int_{{\mathcal{H}}_n} \exp\left(- 
\frac{1}{2}  \trace  (X^2)\right)\cdot \exp\left(\frac{t}{4}
  \trace 
X^4\right)  \frac{d\mu(X)}{N} .
$$
Using the same argument as in Lemma \ref{lma1.2}
we can prove the asymptotic formula
\begin{multline*}
{{\mathcal{A}}}\left(\int_{{\mathcal{H}}_n} \exp\left(- 
\frac{1}{2}  \trace  (X^2)\right)\cdot \exp\left(\frac{t}{4}
  \trace 
X^4\right)  \frac{d\mu(X)}{N}\right)\\
=\sum_{v=0} ^{\infty} \frac{t^{v}}{4!\cdot v!}
\int_{{\mathcal{H}}_n} \exp\left(- 
\frac{1}{2}  \trace  (X^2)\right)\cdot \big(\trace 
X^4\big)^{v}   \frac{d\mu(X)}{N} .
\end{multline*}
We need another matrix $Y=[y_{ij}]\in {\mathcal{H}}_n$ and a
differential operator
$$
\frac{\partial}{\partial Y} =
\left[\frac{\partial}{\partial y_{ij}}\right]
$$
to compute  the asymptotic expansion of the integral.
\begin{lma}\label{lma2.3}
For every $j>0$ and $v >0$, we have
\begin{equation}\label{eq2.2}
\left.\left(\trace  \left(
\frac{\partial}{\partial Y}\right)^j\right)^{v}
e^{\trace  (X^t\cdot Y)}\right|_{Y=0} =
\left(\trace  X^j\right)^{v} .
\end{equation}
\end{lma}
\begin{proof}
Suppose that $Y$ and $X$ are both arbitrary complex
matrices of size $n$. Then for each $j>0$, we have
\begin{multline*}
  \trace 
\left.\left(\frac{\partial}{\partial Y}\right)^{j}
e^{\trace  (X^t\cdot Y)}\right|_{Y=0}\\
=\left.\sum_{i_1,i_2,i_3,\cdots,i_{j} = 1} ^n
\frac{\partial}{\partial y_{i_1i_2}}
\frac{\partial}{\partial y_{i_2i_3}}\cdots
\frac{\partial}{\partial y_{i_{j}i_1}} 
\exp\left(\sum_{k,\ell=1} ^n x_{k\ell}
\cdot y_{k\ell}\right) \right|_{Y=0}\\
=\sum_{i_1,i_2,i_3,\cdots,i_{j} = 1} ^n
x_{i_1i_2} x_{i_2i_3}\cdots x_{i_{j}i_1}=\trace   X^{j} .
\end{multline*}
Repeating it $v$ times, we obtain the desired formula (\ref{eq2.2})
for general complex matrices. Certainly, the formula holds
after changing coordinates:
\begin{equation}\label{eq2.3}
\cases
y_{ij} = u_{ij} + \sqrt{-1} w_{ij}\quad {\text{ for }} i<j\\
y_{ji} = u_{ij} - \sqrt{-1} w_{ij}\quad {\text{ for }} i<j\\
y_{ii} = u_{ii}
\endcases ,
\end{equation}
where $u_{ij}$ and $w_{ij}$ are complex variables. Since
 (\ref{eq2.2})
is an algebraic formula, it holds for an arbitrary field
of characteristic $0$.
In particular, (\ref{eq2.2}) holds 
for real $u_{ij}$ and $w_{ij}$, which
proves the lemma.
\end{proof}
\noindent
Therefore, we have
\begin{align*}
& \int_{{\mathcal{H}}_n} \exp\left(- 
\frac{1}{2}  \trace  (X^2)\right)\cdot \big(\trace 
X^4\big)^{v}   \frac{d\mu(X)}{N}\\
=&\left.\int_{{\mathcal{H}}_n} \exp\left(- 
\frac{1}{2}  \trace  (X^2)\right)\cdot
\left(\trace  \left(
\frac{\partial}{\partial Y}\right)^4\right)^{v}
e^{\trace  (X^t\cdot Y)}\right|_{Y=0}   \frac{d\mu(X)}{N}\\
=&\left.\left(\trace  \left(
\frac{\partial}{\partial Y}\right)^4\right)^{v}
\int_{{\mathcal{H}}_n} \exp\left(- 
\frac{1}{2}  \trace  (X-Y^t)^2\right)\cdot
e^{1/2\trace  (Y^t)^2}\right|_{Y=0}   \frac{d\mu(X)}{N}\\
=&\left.\left(\trace  \left(
\frac{\partial}{\partial Y}\right)^4\right)^{v}
e^{1/2\trace  Y^2}\right|_{Y=0}\\
=&\left.\left(\sum_{i,j,k,\ell}
\frac{\partial}{\partial y_{ij}}
\frac{\partial}{\partial y_{jk}}
\frac{\partial}{\partial y_{k\ell}}
\frac{\partial}{\partial y_{\ell i}}\right)^{v}
\exp\left(\frac{1}{2} \sum_{i,j} y_{ij} y_{ji}\right)
\right|_{Y=0} .
\end{align*}
The only nontrivial contribution of the
differentiation comes from paired derivatives:
$$
\left.\frac{\partial}{\partial y_{ij}}
\frac{\partial}{\partial y_{k\ell}} 
\exp\left(\frac{1}{2} \sum_{i,j}
 y_{ij} y_{ji}\right)\right|_{Y=0}
= \frac{\partial}{\partial y_{ij}}  y_{\ell k}
= \delta_{i\ell}\cdot \delta_{jk} .
$$
If we denote by $\bullet_{ij}$ the differential operator
$\frac{\partial}{\partial y_{ij}}$, then we have a pairing
scheme of $4v$ dots as before, and the pairing
of two dots $\bullet_{ij}$ and $\bullet_{k\ell}$ contributes
$\delta_{i\ell}\cdot \delta_{jk}$. 
Thus
\begin{equation}\label{eq2.4}
\begin{split}
&\left.\left(\trace  \left(
\frac{\partial}{\partial Y}\right)^4\right)^{v}
e^{1/2\trace   Y^2}\right|_{Y=0}\\
&\quad =\sum_{i_1,j_1,k_1,\ell_1=1} ^n \cdots
\sum_{i_{v},j_{v},k_{v},\ell_{v}=1} ^n
\sum_{\substack{{\text{All pairings}}\\
{\text{$P$ of $4v$ dots}}}} \left(
\prod_{\substack{{\text{All paired dots}}\\
(\bullet_{ij},\bullet_{k\ell})   {\text{ in }}
 P}}\delta_{i\ell}\cdot \delta_{jk}\right) .
\end{split}
\end{equation}
A symbolic description of the
contribution of parings is given in Figures~\ref{fig2.85}.

\begin{figure}[htb]
\centerline{\epsfig{file=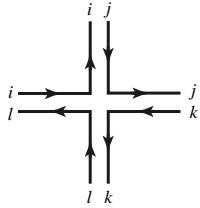, width=1in}}
\caption{Indexed Crossroad}\label{fig2.7} 
\end{figure}

An interpretation of Figure~\ref{fig2.85}
 in terms of Feynman Diagrams
was introduced by 'tHooft \cite{'tHooft}. The set of
four \emph{indexed} dots $\bullet_{ij} \bullet_{jk}
\bullet_{k\ell} \bullet_{\ell i}$ is replaced by a crossroad
(Figure~\ref{fig2.7}).

\begin{figure}[htb]
\centerline{\epsfig{file=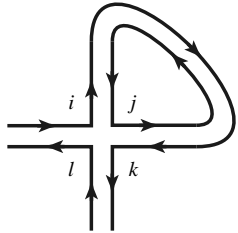, width=1.2in}}
\caption{Crossroad Loop}\label{fig2.8} 
\end{figure}

Since $\bullet_{ij} = \frac{\partial}{\partial y_{ij}}$ is
different from $\bullet_{ji} = \frac{\partial}{\partial y_{ji}}$,
the different roles of the indices are represented by an arrow.
If $\bullet_{ij}$ is connected to $\bullet_{jk}$, then it
gives a contribution of $\delta_{ik}\cdot \delta_{jj}$. 'tHooft
visualized this situation graphically by making a crossroad loop
(Figure~\ref{fig2.8}).

Note that the orientation of the sidewalks of this crossroad loop
is consistent. Thus we obtain a ribbon graph, as we expected.

The passage from the pairing scheme to a ribbon graph has again some
redundancy. In Section~\ref{feynman}, 
the permutation group ${\mathfrak{S}}_j$
appeared for
a  vertex of degree $j$.
 This is due to the fact that a scalar monomial
$x_1x_2\cdots x_j$ is invariant under the ${\mathfrak{S}}_j$-action.
In the case of matrix integrals, a monomial is of type
$\trace (X_1X_2\cdots X_j)$, which is invariant  under the
action of the cyclic group ${\mathbb{Z}}/j{\mathbb{Z}}$,
but not under the full symmetric group
${\mathfrak{S}}_j$. This is the origin
of the appearance of the extra cyclic order of the edges at
each vertex. 

\begin{Def}
Let $P$ be a pairing scheme
of indexed dots and $\Gamma_{rib}$
the corresponding ribbon graph.
Then the group
$$
\prod_j\left({\mathfrak{S}}_{v_j(\Gamma)}
 \rtimes \big({\mathbb{Z}}/j{\mathbb{Z}}\big)^{v_j(\Gamma)}\right)
$$
acts on the set of all pairing schemes.
As before, we define the automorphism group of
a ribbon graph $\Gamma_{rib}$ to be the
isotropy subgroup of the above group that fixes $P$.
As an abstract group, $\Aut(\Gamma_{rib})$ does not
depend on the choice of the pairing scheme of indexed
dots corresponding to $\Gamma_{rib}$.
\end{Def}

\begin{figure}[htb]
\begin{center}
\epsfig{file=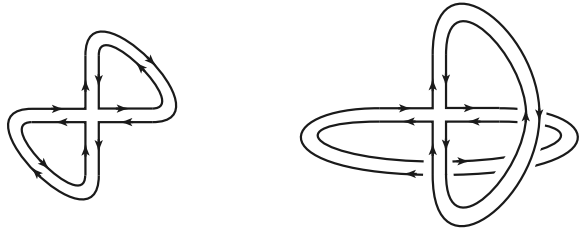, width=3in}
\end{center}
\caption{Degree 4 Ribbon Graphs with 1 Vertex}\label{fig2.9} 
\end{figure}

One more difference between the matrix integral and the
integrals considered in Section~\ref{feynman} 
is the appearance of the
size of matrix in the calculation.
To illustrate this effect, let us continue our
consideration of the degree $4$ case with one vertex:

\begin{equation}
\begin{split}{\label{eq2.45}}
&\frac{1}{4\cdot 1!}  \int_{{\mathcal{H}}_n} e^{-1/2  \trace  (X^2)} 
\trace  (X^4) \frac{d\mu(X)}{N}\\
&= \frac{1}{4}\sum_{i,j,k,\ell = 1} ^n\left(
\delta_{ik}\delta_{jj}\delta_{ki}\delta_{\ell \ell} +
\delta_{i\ell}\delta_{jk}\delta_{ji}\delta_{k\ell} +
\delta_{ii}\delta_{j\ell}\delta_{j\ell}\delta_{kk}\right)\\
&=\frac{1}{4}  (n^3 + n + n^3)
= \frac{1}{2}  n^3 + \frac{1}{4}  n .
\end{split}
\end{equation}
\noindent
As shown in Figure~\ref{fig2.9} there are two degree
$4$ ribbon graphs of order one.
The one on the left has the automorphism
group ${\mathbb{Z}}/2{\mathbb{Z}}$, while the
second has ${\mathbb{Z}}/4{\mathbb{Z}}$. 

We also note that the exponent
of $n$ in (\ref{eq2.45}) is exactly the number of boundary
components of the ribbon graph which is considered as a surface
with boundary. For every $v\ge 1$, we now have
\begin{multline*}
\frac{1}{4^v\cdot v!}  \int_{{\mathcal{H}}_n} e^{-1/2  \trace  (X^2)} 
\big(\trace  (X^4)\big)^v \frac{d\mu(X)}{N}\\
=\sum_{\substack{{\text{degree $4$ ribbon graph}}\\
\Gamma_{rib} {\text{  of order $v$}}}}
 \frac{1}{|\Aut(\Gamma_{rib})|} 
n^{b(\Gamma_{rib})} \in {\mathbb{Q}}[n] .
\end{multline*}

The same argument that we used to prove Theorem~\ref{thm1.5}
works and we have:
\begin{thm}\label{thm2.4}
The asymptotic expansion
of the Hermitian matrix integral (\ref{eq2.1}) is given by
\begin{multline*}
{{\mathcal{A}}}\left(\int_{{\mathcal{H}}_n} \exp\left(- 
\frac{1}{2}  \trace  (X^2)\right) \exp\left(\trace 
\sum_{j = 3} ^{2m}
\frac{t_j }{j} X^j \right)   \frac{d\mu(X)}{N}\right)\\
=\sum_{\substack{{\text{Ribbon graph $\Gamma_{rib}$ with}}\\
{\text{vertices of degree }} 3,4,\cdots, 2m}}
\frac{1}{|\Aut(\Gamma_{rib})|}  n^{b(\Gamma_{rib})}\cdot
\prod_{j=3} ^{2m} {t_j} ^{v_j(\Gamma)} ,
\end{multline*}
where $b(\Gamma_{rib})$ denotes the number of 
boundary components of the ribbon graph  $\Gamma_{rib}$,
and $v_j(\Gamma)$ the number of degree $j$ vertices in
the underlying graph $\Gamma$. 
\end{thm}

\noindent
Here we note that for given values of $v_3(\Gamma), \cdots,
v_{2m}(\Gamma)$, the number of ribbon graphs is finite.
Thus the above asymptotic series belongs to
$\big({\mathbb{Q}}[n]\big)[[t_3,t_4,\cdots,t_{2m}]]$.
The relation between connected ribbon graphs and arbitrary
ribbon graphs are the same as in 
Section~\ref{feynman}.
In particular, since
(\ref{eq1.11}) also holds for ribbon graphs,
application of the logarithm gives us
\begin{thm}\label{thm2.5}
\begin{multline*}
\log {{\mathcal{A}}}\left(\int_{{\mathcal{H}}_n} \exp\left(- 
\frac{1}{2}  \trace  (X^2)\right) \exp\left(\trace 
\sum_{j = 3} ^{2m}
\frac{t_j }{j} X^j \right)   \frac{d\mu(X)}{N}\right)\\
=\sum_{\substack{{\text{Connected ribbon graph $\Gamma_{rib}$}}\\
{\text{with maximum degree }} 2m}}
\frac{1}{|\Aut(\Gamma_{rib})|}  n^{b(\Gamma_{rib})}\cdot
\prod_{j=3} ^{2m} {t_j} ^{v_j(\Gamma)} .
\end{multline*}
\end{thm}
\noindent
This formula is particularly useful, because we are interested
in \emph{connected} Riemann surfaces and only connected ribbon
graphs give rise to connected surfaces. Using Lemma~\ref{lma2.2},
we can rearrange the summation in terms of the
genus of a compact oriented surface and the number of
marked points on it:
\begin{equation}\label{eq2.5}
\begin{split}
&\log {\mathcal{A}}\left(Z_n(t,m)\right)\\
&\qquad=
\sum_{\substack{ g\ge 0,  s>0\\2-2g-s<0}}
\left(
\sum_{\substack{{\text{Connected ribbon graph }}
\Gamma_{rib} \\
{\text{with vertices of degree }} 3, 4, \cdots,2m,\\
\rchi(\Gamma) = 2-2g-s,  b(\Gamma_{rib}) = s}}
\frac{n^{s}}
{|\Aut(\Gamma_{rib})|}\cdot \prod_{j= 3} ^{2m}
t_j ^{v_j(\Gamma)}
\right)  ,
\end{split}
\end{equation}
where
$\rchi(\Gamma)$  denotes the Euler characteristic of the
underlying graph
$\Gamma$. 
Note that Lemma~\ref{lma2.2} implies that
$$
2 - 2g(C(\Gamma_{rib})) -b(\Gamma_{rib})
<0
$$
for every ribbon graph $\Gamma_{rib}$. Let $v(\Gamma)$
and $e(\Gamma)$ be the
total number of vertices and edges
of the graph $\Gamma$, respectively. Then
\begin{equation}\label{eq2.6}
\cases
\rchi(\Gamma) = v(\Gamma) - e(\Gamma)\\
v(\Gamma) = v_3(\Gamma) + v_4(\Gamma) + \cdots +
v_{2m}(\Gamma)\\
e(\Gamma) = \frac{1}{2} \big( 3\cdot v_3(\Gamma) +
4\cdot v_4(\Gamma) + \cdots +
2m\cdot v_{2m}(\Gamma)\big) ,
\endcases
\end{equation}
because the vertices of $\Gamma$ have degree in between
 $3$ and  $2m$.
Thus for every fixed $g$ and $s$, the second
summation of (\ref{eq2.5}) is a finite sum, which again  shows that
(\ref{eq2.5}) is an element of the formal power series ring
$$
\big({\mathbb{Q}}[n]\big)[[t_3, t_4, \cdots, t_{2m}]] .
$$
The number $g$ is of course the genus of
$C(\Gamma_{rib})$. The topological type of the 
ribbon graph $\Gamma_{rib}$ is the same as
the compact surface $C(\Gamma_{rib})$ minus
$b(\Gamma_{rib})$ points. The number of boundary
components becomes
the number of \emph{marked} points of a Riemann
surface in later sections. 

Let
$\big({\mathbb{Q}}[n]\big)[[t_3, t_4,  \cdots]] $
be the formal power series ring
in infinitely
many variables.
The \emph{adic} topology of this ring
is given by the degree
$$
\deg  t_j = j, \qquad j\ge 3
$$
and the ideal ${\mathfrak{I}}_j(t)$ of
$\big({\mathbb{Q}}[n]\big)[[t_3, t_4, \cdots]] $
generated by  polynomials
in $t_3, t_4,\cdots$ of degree
greater than $j$,
with coefficients in ${\mathbb{Q}}[n]$.
We have a natural projection
$$
\pi_j :  \big({\mathbb{Q}}[n]\big)[[t_3, t_4, \cdots]]
\longrightarrow
\big({\mathbb{Q}}[n]\big)[[t_3, t_4, \cdots]]
 \big/ {\mathfrak{I}}_j(t) =
\big({\mathbb{Q}}[n]\big)[[t_3,  \cdots, t_j]]
 \big/ {\mathfrak{I}}_j(t) .
$$
{For} each fixed $j$, the projection image
$$
\pi_j\big( \log{{\mathcal{A}}} \big(Z_n(t,m)\big)
\big)
\in \big({\mathbb{Q}}[n]\big)[[t_3, t_4, \cdots]]
\big/ {\mathfrak{I}}_j(t) =
\big({\mathbb{Q}}[n]\big)[[t_3,  \cdots, t_j]]
 \big/ {\mathfrak{I}}_j(t)
$$
is stable for all $2m \ge j$.  Since
$$
\big({\mathbb{Q}}[n]\big)[[t_3, t_4, \cdots]] =
\lim_{\substack{ \longleftarrow\\ j}}
 \big({\mathbb{Q}}[n]\big)[[t_3, t_4, \cdots]] \big/ {\mathfrak{I}}_j(t)
$$
and
$$
\left\{ \pi_{2m}\big(\log{{\mathcal{A}}}
\big(Z_n(t,m)\big)\big)\right\}_{m\ge 2}
$$
defines an element of the projective system, it gives
a well-defined
formal power series in infinitely many variables.
We denote  it symbolically by
\begin{equation}\label{eq2.7}
\begin{split}
\lim_{m\rightarrow \infty}\log
&{{\mathcal{A}}}\big( Z_n(t,m) \big) \\
&\qquad=
\left\{ \pi_{2m}\big(\log{{\mathcal{A}}}
\big(Z_n(t,m)\big)\big)\right\}_{m\ge 2}
\in \big({\mathbb{Q}}[n]\big)[[t_3,t_4,\cdots]] .
\end{split}
\end{equation}

Going back to the Feynman diagram expansion (\ref{eq2.5}),
we have an equality
\begin{equation}\label{eq2.8}
\begin{split}
\lim_{m\rightarrow \infty}\log
&{{\mathcal{A}}}\big( Z_n(t,m)\big)\\
&\qquad=
\sum_{\substack{ g\ge 0,  s>0\\2-2g-s<0}}
\left(
\sum_{\substack{{\text{Connected ribbon graph }}\Gamma_{rib} \\
{\text{with vertices of degree }} \ge 3,\\
\rchi(\Gamma) = 2-2g-s,  b(\Gamma_{rib}) = s}}
\frac{n^{s}}
{|\Aut(\Gamma_{rib})|}\cdot \prod_{j\ge 3}
t_j ^{v_j(\Gamma)}
\right)
\end{split}
\end{equation}
as an element of
$\big({\mathbb{Q}}[n]\big)[[t_3, t_4,  \cdots]] $.
For each fixed $g$ and $s$, the maximum possible
valency of the ribbon graphs in the second summation is
$4g + 2s -2$.  To see this, let $\Gamma$ be a graph with
the largest possible degree $\ell$. Since the Euler
characteristic of $\Gamma$ is given by
$2-2g-s = v(\Gamma) -e(\Gamma)$,
the degree becomes maximum when
$\Gamma$ has
only one vertex. Thus
$$
2-2g-s = 1- \frac{1}{2}  \ell .
$$
This shows us that the right hand side of (\ref{eq2.8})
does not have any infinite products.

\section{Asymptotic analysis of the Penner model}
\label{asymptotic}
There are no known analytic methods to compute the
matrix integral $Z_n(t,m)$ for general $m$. It is
therefore an amazing observation of Penner  that
at the limit of $m\rightarrow \infty$ a certain
specialization
of $Z_n(t,m)$ is actually computable. In this section
we study the \emph{Penner model} and
calculate its asymptotic expansion analytically.

The specialization Penner considered is the
substitution
\begin{equation}\label{eq3.1}
t_j = -\big(\sqrt z\big)^{j-2}, \quad j=3, 4,
\cdots , 2m
\end{equation}
in the matrix integral $Z_n(t,m)$,
 where $\sqrt z$ is defined for $Re(z) > 0$.
The condition 
$$
\pi/2 + \epsilon <\arg(t_{2m}) <3\pi/2 -\epsilon
$$ for $t_{2m}$ translates
into the condition 
\begin{equation}\label{eq3.2}
|\arg(z)| < \frac{\pi}{2m-2}.
\end{equation}
 Thus we have a holomorphic function
\begin{equation}\label{eq3.3}
\begin{split}
&P_n(z,m)\\
&=
\int_{{\mathcal{H}}_n} \exp\left(- 
\frac{1}{2}  \trace(X^2)\right) \exp\left(
-\sum_{j = 3} ^{2m}
\frac{(\sqrt z)^{j-2} }{j} \trace (X^j )\right)   \frac{d\mu(X)}{N}\\
&=\int_{{\mathcal{H}}_n} \exp\left(
-\sum_{j = 2} ^{2m}
\frac{(\sqrt z)^{j-2} }{j} \trace (X^j )\right)   \frac{d\mu(X)}{N}
\end{split}
\end{equation}
\noindent
defined on the region of the complex plane given
by (\ref{eq3.2}).

\begin{figure}[htb]
\epsfxsize 3in
\centerline{\epsfig{file=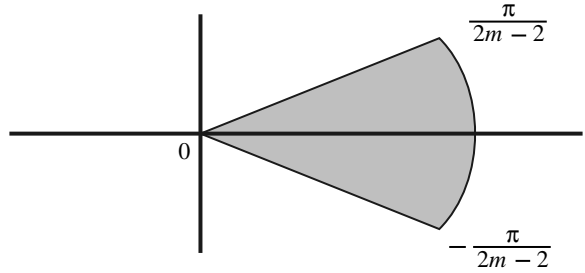, width=3in}}
\caption{Wedge-shape Domain}\label{fig3.1} 
\end{figure}

We note that the  domain (\ref{eq3.2}) still makes sense
 as the
positive real axis when we take the limit
 $m\rightarrow \infty$.
The quantity $N$ is the same normalization
constant as in (\ref{eqnorm}).

  The asymptotic expansion of (\ref{eq3.3}) at $z= 0$
can be calculated by making the same substitution
(\ref{eq3.1}) in Theorem \ref{thm2.4}.  Taking the logarithm, we
obtain
\begin{equation}\label{eq3.4}
\begin{split}
\log{\mathcal{A}}&\left(P_n(z,m)\right) \\
&=\sum_{\substack{ g\ge 0,  s>0\\2-2g-s<0}}
\left(
\sum_{\substack{{\text{Connected ribbon graph }}\Gamma_{rib} \\
{\text{with vertices of degree }} 3, 4, \cdots,2m,\\
\rchi(\Gamma) = 2-2g-s,  b(\Gamma_{rib}) = s}}
\frac{(-1)^{e(\Gamma)}}
{|\Aut(\Gamma_{rib})|}\right)
 n^{s}\cdot (-z)^{2g+s-2}  ,
\end{split}
\end{equation}
where we used (\ref{eq2.6}) to compute
\begin{align*}
\prod_{j= 3} ^{2m}
\left(-(\sqrt z)^{j-2}\right) ^{v_j(\Gamma)}
&= (-1)^{\Sigma_{j=3} ^{2m} v_j(\Gamma)}\cdot
z^{\frac{1}{2} \Sigma_{j=3} ^{2m} jv_j(\Gamma)
-\Sigma_{j=3} ^{2m} v_j(\Gamma)}\\
&=(-1)^{v(\Gamma)} z^{e(\Gamma) - v(\Gamma)}\\
&=(-1)^{e(\Gamma)} (-z)^{-\rchi(\Gamma)} .
\end{align*}
Note that the right hand side of (\ref{eq3.4}) is a well-defined
element of $\big({\mathbb{Q}}[n]\big)[[z]]$.
{For} every $\nu>0$, the terms in
$\log{\mathcal{A}}\left(P_n(z,m)\right)$ of degree less than or
equal to $\nu$ with respect to $z$ are stable for all
$m\ge \nu +1$. Again by the same argument we used in
Section~\ref{matrix}, we can define an element
$$
\lim_{m\rightarrow \infty}
\log{\mathcal{A}}\left(P_n(z,m)\right) \in
\big({\mathbb{Q}}[n]\big)[[z]] .
$$
Thus we have an equality
\begin{equation}\label{eq3.5}
\begin{split}
\lim_{m\rightarrow \infty}\log{{\mathcal{A}}}&\left(
\int_{{\mathcal{H}}_n}  \exp\left(
-\sum_{j = 2} ^{2m}
\frac{(\sqrt z)^{j-2} }{j} \trace (X^j )\right)  \frac{d\mu(X)}{N}
\right)\\
&= \sum_{\substack{ g\ge 0,  s>0\\2-2g-s<0}}
\left(
\sum_{\substack{{\text{Connected ribbon graph }} 
\Gamma_r\\
{\text{ with vertices of degree }}\ge 3,\\
\rchi(\Gamma) = 2-2g-s,  b(\Gamma_{rib}) = s}}
\frac{(-1)^{e(\Gamma)}}
{|\Aut(\Gamma_{rib})|}\right)
 n^{s}\cdot (-z)^{2g+s-2}
\end{split}
\end{equation}
as a well-defined  element of $\big({\mathbb{Q}}[n]\big)[[z]]$.
We recall that in (\ref{eq2.8}) we proved that the number of
ribbon graphs in
the second summation
for  fixed $g$ and $s$ is finite.

Let us now  compute
$\lim_{m\rightarrow \infty}
\log{\mathcal{A}}\left(P_n(z,m)\right)$.
The standard analytic technique to compute the
Hermitian matrix integrals
is the following formula.
Let $f(X)$ be a function on $X\in {\mathcal{H}}_n$ which is
invariant under the conjugation by a unitary matrix
$U\in U(n)$:
$$
f(X) = f(U^{-1}\cdot X\cdot U)
= f(k_0, k_1, \cdots, k_{n-1}) ,
$$
where $k_0, k_1, \cdots, k_{n-1}$ are the eigenvalues
of the Hermitian matrix $X$. If $f(X)$ is integrable on
${\mathcal{H}}_n$ with respect to the measure $d\mu(X)$, then
\begin{equation}\label{eq3.6}
\int_{{\mathcal{H}}_n} f(X)  d\mu(X) =
c(n)\cdot \int_{{\mathbb{R}}^n} f(k_0, k_1, \cdots, k_{n-1})
\Delta(k)^2
dk_0 dk_1\cdots dk_{n-1} ,
\end{equation}
where
\begin{equation}\label{eq3.7}
c(n) =
\frac{\pi^{n(n-1)/2}}{n!\cdot (n-1)!\cdots 2!\cdot 1!} ,
\end{equation}
and
$$
\Delta(k)=\Delta\big(k_0,k_1 \cdots, k_{n-1}\big)
= \det \pmatrix
1 & k_0 & k_0 ^2 & \hdots & k_0 ^{n-1}\\
1 & k_1 & k_1 ^2 & \hdots & k_1 ^{n-1}\\
1 & k_2 & k_2 ^2 & \hdots & k_2 ^{n-1}\\
\vdots & \vdots & \vdots & \ddots & \vdots\\
1 & k_{n-1} & k_{n-1} ^2 & \hdots & k_{n-1} ^{n-1}
\endpmatrix
=\prod_{i>j} (k_i-k_j)
$$
is the Vandermonde determinant. The proof
of (\ref{eq3.6}) goes as follows:

Let $\overset\circ{{\mathcal{H}}}_n$ denote the
open dense subset of ${\mathcal{H}}_n$ consisting of
non-singular Hermitian matrices of size $n$
with $n$ distinct eigenvalues. If $f(X)$ is a
regular integrable function on ${\mathcal{H}}_n$, then
$$
\int_{{\mathcal{H}}_n} f(X)   d\mu(X) =
\int_{\overset\circ{{\mathcal{H}}}_n} f(X)   d\mu(X) .
$$
We denote by $\overset\circ{\mathbb{R}}{}^n$
the space of real diagonal matrices of all
distinct, non-zero eigenvalues. Here again
integration over ${\mathbb{R}}^n$ is equal to
integration over $\overset\circ{{\mathbb{R}}}{}^n$.
Since every Hermitian matrix is diagonalizable
by a unitary matrix, we have a surjective map
$$
U(n)\times \overset\circ{{\mathbb{R}}}{}^n
\owns \left(U,
\bmatrix
k_0\\
&\ddots\\
&&k_{n-1}
\endbmatrix
\right)
\longmapsto
U\cdot
\bmatrix
k_0\\
&\ddots\\
&&k_{n-1}
\endbmatrix
\cdot U^{-1}
\in
\overset\circ{{\mathcal{H}}}_n .
$$
The fiber of this map is the set of all unitary
matrices that are commutative with a generic
real diagonal matrix, which can be identified
with the product of two subgroups
$$
T^n\cdot W_n\subset U(n) ,
$$
where $T^n\subset U(n)$ is
the maximal torus of $U(n)$,
and $W_n\subset U(n)$
the group of permutation matrices
of size $n$. Note that
$$
\dim  U(n) = \dim  {{\mathcal{H}}}_n = n^2, \quad \dim  T^n = n .
$$
Therefore, the induced map
$$
h:U(n)\big/T^n \times \overset\circ{{\mathbb{R}}}{}^n
\longrightarrow \overset\circ{{\mathcal{H}}}_n
$$
is a covering map of degree $|W_n| = n!$.
We need the Jacobian determinant of $h$.
Put
$$
X = \big[ x_{ij}\big]
= U\cdot
\bmatrix
k_0\\
&\ddots\\
&&k_{n-1}
\endbmatrix
\cdot U^{-1}
\in
\overset\circ{{\mathcal{H}}}_n ,
$$
and denote
$$
dX = \big[ dx_{ij}\big] .
$$
Then
\begin{align*}
dX&=dU\cdot
\bmatrix
k_0\\
&\ddots\\
&&k_{n-1}
\endbmatrix
\cdot U^{-1} +
U\cdot
\bmatrix
dk_0\\
&\ddots\\
&&dk_{n-1}
\endbmatrix
\cdot U^{-1}\\
& \quad -U\cdot
\bmatrix
k_0\\
&\ddots\\
&&k_{n-1}
\endbmatrix
\cdot U^{-1}\cdot  dU\cdot U^{-1}\\
&=U\cdot
\bmatrix
dk_0\\
&\ddots\\
&&dk_{n-1}
\endbmatrix
\cdot U^{-1} + \big[dU\cdot U^{-1},X\big]\\
&=U\cdot\left(
\bmatrix
dk_0\\
&\ddots\\
&&dk_{n-1}
\endbmatrix
+ \left[U^{-1}\cdot dU,
\bmatrix
k_0\\
&\ddots\\
&&k_{n-1}
\endbmatrix
\right]\right)
\cdot U^{-1}\\
&=U\cdot\left(
\bmatrix
dk_0\\
&\ddots\\
&&dk_{n-1}
\endbmatrix
+ \left[\big(k_j-k_i\big)d\omega_{ij}\right]
\right)\cdot U^{-1} ,
\end{align*}
where
$$
U^{-1}\cdot dU = \big[d\omega_{ij}\big] ,
$$
which is a skew Hermitian matrix. In terms of the
above expression, we compute
$$
d\mu(X) = dk_0\wedge\cdots\wedge dk_{n-1}
\wedge\left(
\bigwedge_{i<j}(k_j-k_i)^2 Re(d\omega_{ij})
\wedge \Im(d\omega_{ij})\right) .
$$
Thus the integration on
$\overset\circ{{\mathcal{H}}}_n$ is separated
to integration on $U(n)\big/T^n$ and
$\overset\circ{{\mathbb{R}}}{}^n$. Let
$$
c(n) = \frac{1}{n!}\int_{U(n)\big/T^n}
\bigwedge_{i<j} Re(d\omega_{ij})
\wedge \Im(d\omega_{ij}) .
$$
Then we obtain
$$
\int_{\overset\circ{{\mathcal{H}}}_n}
f(X)  d\mu(X) =
c(n) \int_{\overset\circ{{\mathbb{R}}}{}^n}
\Delta(k)^2 f(k_0,\cdots,k_{n-1})
dk_0\cdots dk_{n-1} .
$$
For computation of $c(n)$, we refer to, for example,
Bessis-Itzykson-Zuber \cite{BIZ}.

Using formula (\ref{eq3.6}), we can reduce our
integral to
$$
P_n(z,m) = \frac{c(n)}{N}
\int_{{\mathbb{R}}^n} \Delta(k)^2\prod_{i=0} ^{n-1}\left(
\exp\left(
-\sum_{j = 2} ^{2m}
\frac{(\sqrt z)^{j-2} }{j}  k_i ^j
\right) dk_i\right) .
$$
At this stage, one might want to compute
\begin{align*}
\lim_{m\rightarrow \infty}\exp\left(
-\sum_{j = 2} ^{2m}
\frac{(\sqrt z)^{j-2} }{j}  k_i ^j
\right)
&= \exp
\left(\frac{1}{z}\log(1-\sqrt z k_i ) +\frac{k_i}{\sqrt z}
\right)\\
&= (1-\sqrt z k_i )^{1/z} \cdot e^{k_i/\sqrt z}\\
&= z^{1/z}\cdot e^{1/z}\cdot x^{1/z} \cdot e^{-x} ,
\end{align*}
where
$$
x= \frac{1-\sqrt z k_i}{z} .
$$
Since the above function in $x$ is proportional to the
Laguerre potential, one might expect that
the integral  becomes
computable.  However, such
a substitution requires a very
careful treatment. First of all, we have to justify
the limit $m\rightarrow\infty$ taken \emph{inside} the
 integral over the whole space.
Secondly, the integral with respect
to $k_i$ is for the entire real axis, which translates
to an integral in $x$ again on the entire real line.
Since the Laguerre potential is not integrable for
negative $x$, the above formal computation
cannot be justifiable inside the integral sign.
What should we do, then?

The following is our key idea to compute the Penner
model.
\begin{thm}[\cite{Mulase1995}]\label{thm3.1}
Let ${\mathfrak{I}}_{\nu}(z) = z^{\nu}\cdot {\mathbb{C}}[[z]]$ denote the ideal
of ${\mathbb{C}}[[z]]$ generated by $z^{\nu}$, and
$$
\pi_{\nu}: {\mathbb{C}}[[z]] \longrightarrow {\mathbb{C}}[[z]]\big/
 {\mathfrak{I}}_{\nu}(z)
$$
the natural projection. For an
arbitrary polynomial $p(k)\in
{\mathbb{C}}[k]$, consider the following two asymptotic
series:
$$
a(z,m) =
{{\mathcal{A}}}\left(
\int_{-\infty} ^{\infty} p(k)\cdot  {\text{\rm{exp}}}\left(
-\sum_{j = 2} ^{2m}
\frac{(\sqrt z)^{j-2} }{j}  k ^j
\right) dk\right) \in {\mathbb{C}}[[z]]
$$
as $z\rightarrow +0$ with $|{\text{\rm{arg}}}(z)| <
\frac{\pi}{2m-2}$, and
$$
b(z) = {{\mathcal{A}}}\left( \sqrt z (ez)^{1/z}\int_0 ^\infty
p\left( \frac{1-zx}{\sqrt z}\right) \cdot x^{1/z}
\cdot e^{-x} \cdot dx
\right) \in {\mathbb{C}}[[z]]
$$
as $z\rightarrow +0$ with $z>0$. Then for every $m>2$,
we have
$$
\pi_m\big(a(z,m)\big) = \pi_m\big(b(z)\big)
$$
as an element of ${\mathbb{C}}[[z]]\big/ {\mathfrak{I}}_m(z)$. In other words,
\begin{multline*}
\lim_{m\rightarrow \infty}
{{\mathcal{A}}}\left(
\int_{-\infty} ^{\infty} p(k)\cdot  {\text{\rm{exp}}}\left(
-\sum_{j = 2} ^{2m}
\frac{(\sqrt z)^{j-2} }{j}  k ^j
\right) dk\right)\\*
= {{\mathcal{A}}}\left( \sqrt z (ez)^{1/z}\int_0 ^\infty
p\left( \frac{1-zx}{\sqrt z}\right) \cdot x^{1/z}
\cdot e^{-x} \cdot dx
\right)
\end{multline*}
holds with respect to the
 ${\mathfrak{I}}_m(z)$-adic topology of ${\mathbb{C}}[[z]]$.
\end{thm}

\begin{rem} The above integrals are \emph{never}
equal as holomorphic functions in $z$. The
limit $m\rightarrow\infty$ makes
 sense only for real positive $z$, and
the equality holds only asymptotically.
\end{rem}
\begin{proof}
Putting $y=\sqrt z   k$, we have
\begin{multline*}
\int_{-\infty} ^{\infty} p(k)\cdot  {\text{\rm{exp}}}\left(
-\sum_{j = 2} ^{2m}
\frac{(\sqrt z)^{j-2} }{j}  k ^j
\right) dk
\\
= \frac{1}{\sqrt z}
\int_{-\infty} ^{\infty} p\left(\frac{y}{\sqrt z}\right)\cdot
 {\text{\rm{exp}}}\left(
- \frac{1}{z}\sum_{j = 2} ^{2m}
\frac{y^j }{j}
\right) dy\\
=\int_{-\infty} ^{\infty} d\nu(y,m) ,
\end{multline*}
where
$$
d\nu(y,m) = \frac{1}{\sqrt z}\cdot
 p\left(\frac{y}{\sqrt z}\right)\cdot
 {\text{\rm{exp}}}\left(
- \frac{1}{z}\sum_{j = 2} ^{2m}
\frac{y^j }{j}
\right) dy .
$$
Let us decompose the integral into three pieces:
\begin{equation}\label{eq3.8}
\int_{-\infty} ^{\infty} d\nu(y,m) =
\int_{-\infty} ^{-1} d\nu(y,m) +
\int_{-1} ^{1} d\nu(y,m) +
\int_{1} ^{\infty} d\nu(y,m) .
\end{equation}
Note that the polynomial
$
\sum_{j = 2} ^{2m}
\frac{y^j }{j}
$
of degree $2m$ takes positive values on the intervals
$(-\infty, -1]$ and $[1, \infty)$.
Since $p(k)$ is a
polynomial, it is obvious that the asymptotic
expansion of the first and the third integrals of
the right hand side of (\ref{eq3.8})
for $z\rightarrow +0$ with $z>0$ is the $0$-series.
Therefore,  we have
$$
\int_{-\infty} ^{\infty} d\nu(y,m)
\overset{A}\equiv
\int_{-1} ^{1} d\nu(y,m) .
$$
On the interval $[-1,1]$, if we fix a $z$ such that
$Re(z)>0$, then
the convergence
$$
\lim_{m\rightarrow \infty}  {\text{\rm{exp}}}\left(
- \frac{1}{z}\sum_{j = 2} ^{2m}
\frac{y^j }{j}
\right)
= (1-y)^{1/z}\cdot e^{y/z}
$$
is absolute and uniform with respect to $y$. Thus,
for a new variable $t = 1 - y$, we have
\begin{align*}
&\lim_{m\rightarrow \infty} \int_{-1} ^{1} d\nu(y,m)
\\ 
&= \frac{1}{\sqrt z}
\int_{-1} ^{1} p\left(\frac{y}{\sqrt z}\right) 
(1-y)^{1/z}  e^{y/z}   dy \\
&=  \frac{1}{\sqrt z}  e^{1/z}  
\int_0 ^2 p\left( \frac{1-t}{\sqrt z}\right)   t^{1/z} 
e^{-t/z}   dt\\
&= \frac{1}{\sqrt z}  e^{1/z}  
\int_0 ^{\infty} p\left( \frac{1-t}{\sqrt z}\right)   t^{1/z} 
e^{-t/z}   dt 
- \frac{1}{\sqrt z}  e^{1/z}  
\int_2 ^{\infty} p\left( \frac{1-t}{\sqrt z}\right)   t^{1/z} 
e^{-t/z}   dt .
\end{align*}
This last integral is
$$
\frac{1}{\sqrt z}  e^{1/z}  
\int_2 ^{\infty} p\left( \frac{1-t}{\sqrt z}\right)   t^{1/z} 
e^{-t/z}   dt =
\frac{1}{\sqrt z} 
\int_2 ^{\infty} p\left( \frac{1-t}{\sqrt z}\right) 
e^{(1+\log t -t)/z}    dt .
$$
Since $1 + \log t -t < 0$ for $t\ge 2$,
the asymptotic expansion
of this integral as $z\rightarrow +0$ with $z>0$ is the $0$-series.
Therefore, since the integrals do not depend on the
integration variables, we have
\begin{multline*}
\lim_{m\rightarrow \infty}
{{\mathcal{A}}}\left( \int_{-\infty} ^{\infty} p(k)\cdot
 {\text{\rm{exp}}}\left(
-\sum_{j = 2} ^{2m}
\frac{(\sqrt z)^{j-2} }{j}  k ^j
\right) dk \right)\\*
= {{\mathcal{A}}}\left(\frac{1}{\sqrt z}  e^{1/z}
\int_0 ^{\infty} p\left( \frac{1-t}{\sqrt z}\right)   t^{1/z} 
e^{-t/z}   dt \right)\\*
=  {{\mathcal{A}}}\left(\sqrt z  e^{1/z}   z^{1/z}
\int_0 ^{\infty} p\left( \frac{1-zx}{\sqrt z}\right)   x^{1/z} 
e^{-x}   dx \right)
\end{multline*}
as a formal power series in $z$.
This completes the proof of Theorem.
\end{proof}
By applying  Theorem \ref{thm3.1} for each $k_i$, we obtain
\begin{equation}\label{eq3.9}
\begin{split}
&\lim_{m\rightarrow \infty}
{{\mathcal{A}}}\left( \int_{{\mathbb{R}}^n} \Delta(k)^2\cdot
\prod_{i=0} ^{n-1} {\text{\rm{exp}}}\left(
-\sum_{j = 2} ^{2m}
\frac{(\sqrt z)^{j-2} }{j}  k_i ^j
\right) dk_i \right)\\
&=  {{\mathcal{A}}}\left( \left(\sqrt z  e^{1/z}   z^{1/z}\right)^n
\int_0 ^{\infty} \cdots \int_0 ^{\infty}
\Delta\left( \frac{1-zx}{\sqrt z}\right)^2\cdot
 \prod_{i=0} ^{n-1}
 x_i ^{1/z} 
e^{-x_i}   dx_i\right)\\
&=  {{\mathcal{A}}}\left( \left(\sqrt z  e^{1/z}   z^{1/z}\right)^n
z^{\frac{n(n-1)}{2}}\int_0 ^{\infty} \cdots \int_0 ^{\infty}
\Delta( x)^2\cdot
 \prod_{i=0} ^{n-1}
 x_i ^{1/z} 
e^{-x_i}   dx_i\right) ,
\end{split}
\end{equation}
where we used the multilinear property of the
Vandermonde determinant.
We can use the standard technique of orthogonal
polynomials to compute the above integral.
Let $p_j(x)$ be a monic orthogonal polynomial
in $x$ of degree
$j$ with respect to the measure
$$
d\lambda(x) = x^{1/z} e^{-x} dx
$$
defined on $K = [0,\infty) $ for a positive $z >0$:
$$
\int_{K} p_i(x) p_j(x) d\lambda(x) =
\delta_{ij} \parallel p_j(x)
\parallel^2 .
$$
Because of the multilinearity of the determinant, we have
once again
$$
\Delta (x) = \det\left(x_i ^j\right)
= \det \left(p_j(x_i)\right) .
$$
Therefore,
\begin{equation}\label{eq3.10}
\begin{split}
  &\int_{K^n} \Delta(x)^2d\lambda(x_0)\cdots
d\lambda(x_{n-1})\\
&=  \int_{K^n} \det \left(p_j(x_i)\right)
\det \left(p_j(x_i)d\lambda(x_i)\right)\\
&=  \int_{K^n} \sum_{\sigma\in {{\mathfrak{S}}}_n}
\sum_{\tau\in {{\mathfrak{S}}}_n}
{\text{sign}}(\sigma)  {\text{sign}}(\tau)
\prod_{i=0} ^{n-1} p_{\sigma(i)}(x_i)
\prod_{i=0} ^{n-1} p_{\tau(i)}(x_i)d\lambda(x_i)\\
&=  \sum_{\sigma\in {{\mathfrak{S}}}_n}
\sum_{\tau\in {{\mathfrak{S}}}_n}
{\text{sign}}(\sigma)  {\text{sign}}(\tau)
\prod_{i=0} ^{n-1}\int_{K}
p_{\sigma(i)}(x) p_{\tau(i)}(x)d\lambda(x)\\
&=  \sum_{\sigma\in {{\mathfrak{S}}}_n}
{\text{sign}}(\sigma)^2\prod_{i=0} ^{n-1}\int_{K}
p_{\sigma(i)}(x) p_{\sigma(i)}(x)d\lambda(x)\\
&=  n! \prod_{i=0} ^{n-1} \parallel p_i (x) \parallel^2 .
\end{split}
\end{equation}
{For} a real number $z >0$, the \emph{Laguerre} polynomial
$$
L_m ^{1/z} (x) = \sum_{j=0} ^m
\binom{m+ 1/z}{m-j} \frac{(-1)^j}{j!} x^j
= \frac{(-1)^m}{m!} x^m + \cdots
$$
of degree $m$ satisfies the orthogonality condition
\begin{equation}\label{eq3.11}
\int_0 ^\infty L_i ^{1/z}(x) L_j ^{1/z}(x)  e^{-x} x^{1/z}
dx
= \delta_{ij} \frac{(j+ 1/z)!}{j!} .
\end{equation}
Thus we can use
\begin{equation}\label{eq3.12}
p_i(x) = (-1)^i\cdot i! \cdot L_i ^{1/z}(x)
\end{equation}
for the computation. From \eqref{eq3.9}--\eqref{eq3.12}, we have
\begin{equation}\label{eq3.13}
\begin{split}
&\lim_{m\rightarrow \infty}
{{\mathcal{A}}}\left( \int_{{\mathbb{R}}^n} \Delta(k)^2\cdot
\prod_{i=0} ^{n-1} {\text{\rm{exp}}}\left(
-\sum_{j = 2} ^{2m}
\frac{(\sqrt z)^{j-2} }{j}  k_i ^j
\right) dk_i \right)\\
&\quad=  {{\mathcal{A}}}\left( \left(\sqrt z  e^{1/z}   z^{1/z}\right)^n
z^{\frac{n(n-1)}{2}}n! \prod_{i=0} ^{n-1} i!\cdot
\left( i + \frac{1}{z}\right)!\right) \\
&\qquad=  {{\mathcal{A}}}\left( (ez)^{\frac{n}{z}} \cdot
z^{\frac{n^2}{2}}\cdot n! \prod_{i=0} ^{n-1} i!\cdot
\left( -1 + \frac{1}{z}\right)!\cdot
\left( i + \frac{1}{z}\right)^{n-i}\right).
\end{split}
\end{equation}
Applying \eqref{eq3.6} and \eqref{eq3.13} to \eqref{eq3.5}, we conclude
\begin{equation}\label{eq3.14}
\begin{split}
&\lim_{m\rightarrow \infty}\log{{\mathcal{A}}}\left( \frac{1}{N}
\int_{{\mathcal{H}}_n}  \exp\left(
-\sum_{j = 2} ^{2m}
\frac{(\sqrt z)^{j-2} }{j} \trace (X^j )\right)   d\mu(X)
\right)\\
&=    \log{{\mathcal{A}}}\left( \frac{1}{N}\cdot
\pi^{\frac{n(n-1)}{2}}\cdot
 (ez)^{\frac{n}{z}} \cdot
z^{\frac{n^2}{2}}\cdot \prod_{i=0} ^{n-1}
\left( -1 + \frac{1}{z}\right)!\cdot
\left( i + \frac{1}{z}\right)^{n-i}\right)\\
&=    \log{{\mathcal{A}}}\left( \frac{1}{N}\cdot
\pi^{\frac{n(n-1)}{2}}\cdot
 (ez)^{\frac{n}{z}} \cdot
z^{\frac{n^2}{2}}\cdot \left(\Gamma\left(\frac{1}{z}
\right)\right)^n\cdot
\prod_{i=0} ^{n-1}
\left( i + \frac{1}{z}\right)^{n-i}\right)\\
&=   \text{const} + \frac{n}{z} +
\frac{n}{z}\log z + \frac{n^2}{2}\log z
+ n\log{{\mathcal{A}}}\left(\Gamma\left(\frac{1}{z}\right) \right)\\
&\qquad+ \sum_{i=0} ^{n-1} (n-i) \log\frac{1+iz}{z}\\
&=   \text{const} + \frac{n}{z} +
\frac{n}{z}\log z - \frac{n}{2}\log z
+ n\log{{\mathcal{A}}}\left(\Gamma\left(\frac{1}{z}\right) \right)\\
&\qquad + \sum_{r=1} ^\infty \frac{(-1)^{r-1}}{r}
\left(\sum_{i=0} ^{n-1} (n-i)i^r\right) z^r .
\end{split}
\end{equation}
Let us recall
Stirling's formula:
\begin{equation}\label{eq3.15}
\log{{\mathcal{A}}}\left(\Gamma\left(\frac{1}{z}\right)\right)
= -\frac{1}{z}\log z -\frac{1}{z} + \frac{1}{2}\log z +
\sum_{r = 1} ^\infty \frac{b_{2r}}{2r(2r-1)}
z^{2r-1} + \text{const} ,
\end{equation}
where $b_r$ is the Bernoulli number defined by
$$
\frac{x}{e^x -1} =
\sum_{r=0} ^\infty \frac{b_r}{r!} x^{r} .
$$
We are not interested in the constant term
(the term independent of $z$) of \eqref{eq3.15}
because the asymptotic series in question, \eqref{eq3.5}, has
no constant term.  We can see that substitution of \eqref{eq3.15}
in \eqref{eq3.14} eliminates all the logarithmic terms as desired:
\begin{multline*}
\lim_{m\rightarrow \infty}\log{{\mathcal{A}}}\left( \frac{1}{N}
\int_{{\mathcal{H}}_n}  \exp\left(
-\sum_{j = 2} ^{2m}
\frac{(\sqrt z)^{j-2} }{j} \trace (X^j )\right) d\mu(X)
\right)\\
=    \sum_{r = 1} ^\infty \frac{b_{2r}}{2r(2r-1)}
 \cdot n\cdot z^{2r-1}
+ \sum_{r=1} ^\infty \frac{(-1)^{r-1}}{r}
\left(\sum_{i=0} ^{n-1} (n-i)i^r\right) z^r .
\end{multline*}
Let
$$
\phi_r(x) = \sum_{q=0} ^{r-1} \binom{r}{q} b_q x^{r - q}
$$
denote the Bernoulli polynomial. Then we have
$$
\sum_{i=1} ^{n-1} i^{r} = \frac{\phi_{r+1}(n)}{r+1}   .
$$
Thus for $r>0$,
\begin{multline*}
\sum_{i=0} ^{n-1} (n-i)i^{r} = \frac{n\phi_{r+1}(n)}{r+1}
- \frac{\phi_{r+2}(n)}{r+2}\\
= \sum_{q=0} ^r \frac{1}{r+1} \binom{r+1}{q}
b_q\cdot n^{r+2-q}
-\sum_{q=0} ^{r+1} \frac{1}{r+2} \binom{r+2}{q}
b_q\cdot n^{r+2-q}\\
= \sum_{q=0} ^r \frac{r!  (1-q)}{q!  (r+2-q)!} 
b_q\cdot n^{r+2-q} - b_{r+1}\cdot n  .
\end{multline*}
Therefore, we have
\begin{equation}\label{eq3.16}
\begin{split}
&\sum_{r = 1} ^\infty \frac{b_{2r}}{2r(2r-1)}
 \cdot n\cdot z^{2r-1}
+ \sum_{r=1} ^\infty \frac{(-1)^{r-1}}{r}
\left(\sum_{i=0} ^{n-1} (n-i)i^r\right) z^r\\
&=   -\sum_{r = 1} ^\infty \frac{1}{2r} 
b_{2r} \cdot n\cdot z^{2r-1}
+ \sum_{r=1} ^\infty\sum_{q=0} ^r
(-1)^{r}\frac{(r-1)!  (q-1)}{q!  (r+2-q)!}
  b_q\cdot n^{r+2-q}\cdot z^r\\
&=   -\sum_{r = 1} ^\infty \frac{1}{2r} 
b_{2r} \cdot n\cdot z^{2r-1}
+ \sum_{r=1} ^\infty (-1)^{r-1} \frac{1}{r(r+1)(r+2)} 
n^{r+2}\cdot z^r
\\
& \qquad +\sum_{r=2} ^\infty\sum_{q=1} ^{[r/2]}
(-1)^{r}\frac{(r-1)!  (2q-1)}{(2q)!  (r+2-2q)!}
  b_{2q}\cdot n^{r+2-2q}\cdot z^r .
\end{split}
\end{equation}
It is time  to switch the summation indices $r$ and $q$ to
$g$ and $s$ as in (\ref{eq3.5}). The first sum of the third
line of (\ref{eq3.16}) is the
case when we specify a single point on a Riemann surface
of arbitrary genus $g = r$. The second sum is for
genus 0 case with more than two points specified. So
we use $s = r+2$ for the number of points.  In the third
sum, $q = g \ge 0$ is the genus and $r + 2 -2q = s \ge 2$
is the number of points. Thus (\ref{eq3.16}) is equal to
\begin{equation}\label{eq3.17}
\begin{split}
 &\sum_{g=1} ^\infty \zeta(1-2g)\cdot n\cdot z^{2g-1} +
\sum_{s=3} ^\infty (-1)^{s-1} \frac{1}{s(s-1)(s-2)}  n^s\cdot
z^{s-2}\\
& + \sum_{g=1} ^\infty \sum_{s=2} ^\infty
(-1)^{s-1} \frac{(2g+s-3)!}{(2g-2)!  s!} 
\zeta(1-2g)\cdot n^s\cdot z^{-2+2g+s} ,
\end{split}
\end{equation}
where we used Euler's formula
$$
\zeta(1-2g) = -\frac{b_{2g}}{2g} ,
$$
and the fact that $b_0 = 1$ and $b_{2q + 1} = 0$ for
$q \ge 1$. Note that the first two summations of
(\ref{eq3.17}) are actually the special cases of the third summation
corresponding to $s = 1$ and $g=0$.
Thus we have established:

\begin{thm}\label{thm3.2}
\begin{multline*}
   \lim_{m\rightarrow \infty}\log{{\mathcal{A}}}\left( \frac{1}{N}
\int_{{\mathcal{H}}_n}  \exp\left(
-\sum_{j = 2} ^{2m}
\frac{(\sqrt z)^{j-2} }{j} \trace (X^j )\right)  d\mu(X)
\right)\\
=  -\sum_{\substack{ g\ge 0,  s>0\\2-2g-s<0}}
\frac{(2g+s-3)!(2g)(2g-1)}{(2g)!  s!} 
\zeta(1-2g)\cdot n^s\cdot (-z)^{-2+2g+s} .
\end{multline*}
Since the asymptotic expansion is unique,
from \eqref{eq3.5} we obtain
\begin{equation}\label{penner}
\sum_{\substack{{\text{Connected ribbon graph }}
\Gamma_{rib} \\
{\text{ with vertices of degree }}\ge 3,\\
\rchi(\Gamma) = 2-2g-s,  b(\Gamma_{rib}) = s}}
\frac{(-1)^{e(\Gamma)}}
{|\Aut(\Gamma_{rib})|}
= -  \frac{(2g+s-3)!(2g)(2g-1)}{(2g)!  s!} 
\zeta(1-2g)
\end{equation}
for every $g\ge 0$ and $s>0$ subject to
$2-2g-s<0$.
\end{thm}
\begin{rem}
If we have taken into
account  the values of $c(n)$ and $N$ in the
above computation, then we will
see that all the constant
terms appearing in the
computation automatically cancel out.
\end{rem}

Let us examine a couple of examples.
\begin{ex}\label{gs03}
The simplest case is $g=0$ and $s=3$. 
The underlying graph $\Gamma$ of a ribbon
graph $\Gamma_{rib}$,
whose topological type is $S^2$ minus
three 
points, should satisfy
\begin{align}
&\rchi(\Gamma) = v(\Gamma) - e(\Gamma)=2-2g -s =-1
\qquad{\text{and}}\label{euler} \\
&3v(\Gamma) \le 2e(\Gamma). \label{valency}
\end{align}
Eqn.(\ref{euler}) gives the Euler characteristic of 
a tri-punctured sphere, and Eqn.(\ref{valency}) 
states that every vertex of $\Gamma$
has degree at least 3. It follows from these 
conditions that
$$
e(\Gamma) \le 3.
$$
There are only three graphs in this case, as
shown in Figure~\ref{figgs03}.

\begin{figure}[htb]
\centerline{\epsfig{file=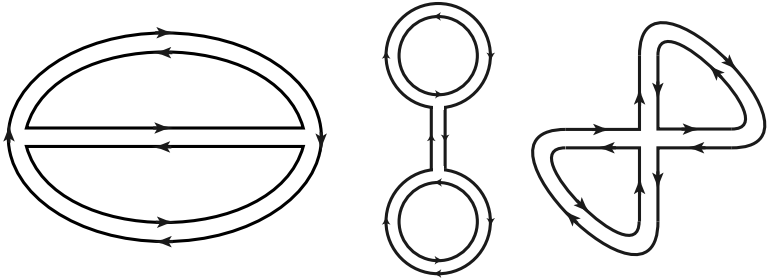, width=4in}}
\caption{Ribbon Graphs for $g=0, s=3$}\label{figgs03} 
\end{figure}

The automorphism groups of  these ribbon
graphs are $\mathfrak{S} _2\rtimes \mathbb{Z}/3\mathbb{Z}
=\mathfrak{S} _3$, $\mathbb{Z}/2\mathbb{Z}$, 
and again $\mathbb{Z}/2\mathbb{Z}$, respectively. 
Thus the left hand side of (\ref{penner}) is 
$$
\frac{(-1)^3}{3!} + \frac{(-1)^3}{2} + \frac{(-1)^2}{2} = -\frac{1}{6}.
$$
The right hand side  is
coming from the term  $n^3 (-z)^1$ of the 
second summation in
(\ref{eq3.17}). The value is, of course, 
$$
-\frac{1}{3(3-1)(3-2)}=-\frac{1}{6}.
$$
It can be also computed from (\ref{penner}):
$$
-\frac{(2g+s-3)!(2g)(2g-1)}{(2g)!s!}\zeta(1-2g)
=\frac{(2g)(2g-1)b_{2g}}{(2g)!3!(2g)}
=-\frac{1}{6}.
$$
\end{ex}

\begin{ex}\label{gs11}
The next simple case is $g=s=1$. Since the 
Euler characteristic condition is  the same
as in Example~\ref{gs03}, the only possibilities 
are again graphs with 1 vertex and 2 edges or 2 vertices and
3 edges. There are two ribbon graphs satisfying 
the conditions: Figure~\ref{fig2.5} and 
the graph on the right in Figure~\ref{fig2.9}.
The first one has 
$\mathfrak{S} _2\times \mathbb{Z}/3\mathbb{Z}$ as its
automorphism group, which happens to be 
a degenerate case of the semi-direct product.
The automorphism group of the second graph is
$\mathbb{Z}/4\mathbb{Z}$, as noted in 
Section~\ref{matrix}.
Thus we have
$$
\frac{(-1)^3}{6}+\frac{(-1)^2}{4} = \frac{1}{12}
= -\zeta(-1).
$$
\end{ex}

\section{KP equations and matrix integrals}
\label{kp}

There are no analytic methods of evaluating the
Hermitian matrix integral
$$
Z_n(t,m) = \int_{{\mathcal{H}}_n} \exp\left(- 
\frac{1}{2}  \trace  (X^2)\right) \exp\left(\trace 
\sum_{j = 3} ^{2m}
\frac{t_j }{j} X^j \right)   \frac{d\mu(X)}{N} .
$$
However, there is an interesting fact about this integral:
\emph{it
satisfies the system
of the KP equations}. In this section we
give a proof of this fact.

To investigate the
most general case, we define
\begin{equation}\label{eq4.1}
Z_n(t,m, \phi) = \int_{{\mathcal{H}}_n}  \exp\left(\trace 
\sum_{j = 1} ^{2m}
\frac{t_j }{j} X^j \right) \phi(X)    \frac{d\mu(X)}{N} ,
\end{equation}
where $\phi(X)$ is a $U(n)$-invariant function on
${\mathcal{H}}_n$  which is determined by $n$ functions $\phi_0(k),
\cdots, \phi_{n-1}(k)$ in
one variable in the following manner:
\begin{equation}\label{eq4.2}
\phi(X) = \phi(k_0, k_1, \cdots k_{n-1}) =
\frac{ \det
\pmatrix
\phi_0(k_0) & \phi_1(k_0) & \hdots & \phi_{n-1}(k_0)\\
\phi_0(k_1) & \phi_1(k_1) & \hdots & \phi_{n-1}(k_1)\\
\vdots & \vdots & \ddots & \vdots\\
\phi_0(k_{n-1}) & \phi_1(k_{n-1}) & \hdots &
\phi_{n-1}(k_{n-1})
\endpmatrix}{\Delta\big(k_0,k_1 \cdots, k_{n-1}\big)} ,
\end{equation}
where $k_0,\cdots, k_{n-1}$ are eigenvalues of $X$.
 Unlike (\ref{eq2.1}), we allow
terms containing
 $t_1X$ and $t_2X^2$  in the integral (\ref{eq4.1}).
Using (\ref{eq3.6}),
we have
\begin{multline*}
Z_n(t,m,\phi) 
= \int_{{\mathcal{H}}_n}
\exp\left(\trace\sum_{\alpha=1} ^{2m}
 \frac{t_\alpha}{\alpha} X^\alpha\right)\phi(X)\cdot
\frac{d\mu(X)}{N}\\
= \frac{c(n)}{N}\int_{{\mathbb{R}}^n} \exp\left(
\sum_{i=0} ^{n-1}\sum_{\alpha = 1} ^{2m}
\frac{t_\alpha}{\alpha}  k_i ^\alpha
 \right)\Delta(k_0,\cdots,k_{n-1})
\det\left(\phi_j(k_i)\right) dk_0\cdots dk_{n-1} .
\end{multline*}
Here we need a simple formula.
Let $\phi_0(k), \cdots, \phi_{n-1}(k)$ and $\psi_0(k),\cdots,
\psi_{n-1}(k)$ be $2n$ arbitrary functions in $k$. Then
\begin{equation}\label{eq4.3}
\det\left[\phi_i(k_\ell)\right]\cdot
\det\left[\psi_j(k_\ell)\right] =
\sum_{\sigma\in {\mathfrak{S}}_n}
\det\left[\phi_i(k_{\sigma(j)})\cdot
\psi_j(k_{\sigma(j)})\right] , 
\end{equation}
where $\sigma$ runs over all permutations of ${\mathfrak{S}}_n$.
To prove (\ref{eq4.3}), we calculate the left hand side by the usual
product formula of the determinant.
Then it becomes a summation of $n^n$
terms. Because of the multilinearity of the determinants, only
$n!$ of these terms are nonzero. Rearranging the $n!$ terms, we
obtain the above formula. Using  this formula
for  $\psi_j(k) = k^j$, we obtain
\begin{equation*}
\begin{split}
&Z_n(t,m,\phi)\\
&= \frac{c(n)}{N}\int_{{\mathbb{R}}^n} \exp\left(
\sum_{i=0} ^{n-1}\sum_{\alpha = 1} ^{2m}
\frac{t_\alpha}{\alpha}  k_i ^\alpha
 \right) \sum_{\sigma\in {\mathfrak{S}}_n}
\det\left(\phi_j(k_{\sigma(i)})k_{\sigma(i)} ^i
\right) dk_0\cdots dk_{n-1}\\
&= \frac{c(n)}{N}\int_{{\mathbb{R}}^n}
\sum_{\sigma\in {\mathfrak{S}}_n}
\exp\left(\sum_{i=0} ^{n-1}
\sum_{\alpha = 1} ^{2m}
\frac{t_\alpha}{\alpha}  k_{\sigma(i)} ^\alpha
 \right)
\det\left(\phi_j(k_{\sigma(i)})k_{\sigma(i)} ^i
\right)  dk_0\cdots dk_{n-1}\\
&= \frac{c(n)}{N}\sum_{\sigma\in {\mathfrak{S}}_n}
\det\left(\int_{{\mathbb{R}}^n} \exp{\left(
\sum_{\alpha = 1} ^{2m}
\frac{t_\alpha}{\alpha}   k_{\sigma(i)} ^\alpha
 \right)} \phi_j(k_{\sigma(i)})k_{\sigma(i)} ^i
\right) dk_0\cdots dk_{n-1}\\
&= n!\cdot \frac{c(n)}{N}
\det\left(\int_{-\infty} ^\infty \exp{\left(
\sum_{\alpha = 1} ^{2m}
\frac{t_\alpha}{\alpha}   k_i ^\alpha
 \right)} \phi_j(k_i)  k_i ^i  dk_i
\right) \\
&= n!\cdot \frac{c(n)}{N}
\det\left(\int_{-\infty} ^\infty \exp{\left(
\sum_{\alpha = 1} ^{2m}
\frac{t_\alpha}{\alpha}   k^\alpha
 \right)} \phi_j(k)  k^i  dk
\right)  .
\end{split}
\end{equation*}
The above computation makes sense as a
complex analytic function
in
$$
(t_1, \cdots, t_{2m-1}, t_{2m})
\in {\mathbb{C}}^{2m-1}\times \{t_{2m}\in{\mathbb{C}} | 
Re(t_{2m})<0\} ,
$$
on which the integral converges, provided that
$|\phi_j(k)|$ grows slower than $\exp(k^{2m})$.
To compare our $t_j$'s with the standard time variables in the KP
theory, let us set
$$
T_{\alpha} = \frac{t_\alpha}{\alpha} .
$$
Now we use the formula
\begin{equation}\label{eq4.4}
\exp\left(\sum_{\alpha = 1} ^{2m} T_\alpha   k^\alpha \right)
= \sum_{r = 0} ^\infty p_r(T)   k^r ,
\end{equation}
where
\begin{equation}\label{eq4.5}
p_r(T) = \sum_{n_1 +2n_2 +3n_3+ \cdots
+ (2m)n_{2m}=r}
\frac{T_1 ^{n_1}\cdot T_2 ^{n_2}\cdot T_3 ^{n_3}\cdots
T_{2m} ^{n_{2m}}}{n_1 !
\cdot n_2 !\cdot n_3 !\cdots n_{2m}!}
\end{equation}
is a weighted homogeneous
polynomial in ${\mathbb{Q}}[T_1, \cdots, T_{2m}]$
of  degree $r$. The relation (\ref{eq4.4}) holds as an entire
function in $T_1,\cdots, T_{2m}$ and $k$.
Note that we have encountered this formula already as
(\ref{eq1.8}).
{From} (\ref{eq4.4}),  we have

\begin{multline*}
Z_n(t,m,\phi)
= n!\cdot \frac{c(n)}{N}
\det\left(\int_{-\infty} ^\infty
\sum_{r = 0} ^\infty p_r(T)   k^r 
  \phi_j(k)  k^i  dk
\right)\\
= n!\cdot \frac{c(n)}{N}
\det\left(\int_{-\infty} ^\infty
\sum_{r = 0} ^\infty p_{r-i}(T)   k^r 
  \phi_j(k)  dk
\right) ,
\end{multline*}
where we define $p_r(T) = 0$ for $r<0$.
\begin{lma}\label{lma4.1}
Let $\phi_j(k)$, $j=0, \cdots, n-1$,
 be a function defined on ${\mathbb{R}}$
such that
$$
\int_{-\infty} ^{\infty} k^r   \phi_j(k)  dk
$$
exists for all $r\ge 0$.  Then as a holomorphic
function defined for $Re(t_{2m})<0$,  we have
$$
{{\mathcal{A}}}\left(\int_{-\infty} ^\infty \exp{\left(
\sum_{\alpha = 1} ^{2m}
\frac{t_\alpha}{\alpha}   k^\alpha
 \right)} \phi_j(k)  k^i  dk\right)
= \sum_{r = 0} ^\infty p_{r-i}(T)
\int_{-\infty} ^\infty
  k^r  \phi_j(k)  dk
$$
as $t_{2m}\rightarrow 0$.
\end{lma}
\begin{proof} The argument is the same as the one
we used in Section~\ref{feynman}. We choose a fixed $t_{2m}$
so that $Re(t_{2m})<0$. Because of the
uniform convergence of the power series
expansion of the integrand, we can interchange
the integral and the infinite sums for
$\alpha = 1, \cdots, 2m-1$. Using (\ref{eq1.7}),
(\ref{eq4.4}) and (\ref{eq4.5}), we have
\begin{equation*}
\begin{split}
 &{{\mathcal{A}}}\left(
\int_{-\infty} ^\infty \exp{\left(
\sum_{\alpha = 1} ^{2m}
T_\alpha   k^\alpha
 \right)} \phi_j(k)  k^i  dk\right)\\
&=\sum_{n_1=0} ^\infty
\frac{{T_1}^{n_1}}{n_1!}
\cdots
\sum_{n_{2m}=0} ^\infty
\frac{{T_{2m}}^{n_{2m}}}
{n_{2m}!}
\int_{-\infty} ^\infty
 k^{i+n_1+2n_2+\cdots+(2m)n_{2m}}
  \phi_j(k)  dk\\
&=\sum_{r = 0} ^\infty p_{r}(T)
\int_{-\infty} ^\infty
  k^{i+r}  \phi_j(k)  dk .
\end{split}
\end{equation*}
\end{proof}
\noindent
Thus we have established
\begin{equation}\label{eq4.6}
\begin{split}
{{\mathcal{A}}}\left(Z_n(t,m,\phi)\right)
&= n!\cdot \frac{c(n)}{N}
\det\left(\sum_{r = 0} ^\infty p_{r-i}(T)
\int_{-\infty} ^\infty
  k^r  \phi_j(k)  dk
\right)\\
&= \det\left(\sum_{r = 0} ^\infty p_{r-i}(T)
\xi_{r j}\right) ,
\end{split}
\end{equation}
where
$$
\xi_{r j} = n!\cdot \frac{c(n)}{N} \int_{-\infty} ^\infty
  k^r \phi_j(k)  dk  .
$$
We recall that the determinant in (\ref{eq4.6}) is an $n\times n$
determinant. Sato \cite{Sato} proved that any size
determinant of the form
\begin{equation}\label{eq4.7}
\det\left(\sum_{r = 0} ^\infty p_{r-i}(T)
\xi_{r j}\right)
\end{equation}
satisfies the Hirota bilinear form of
 the KP equations. He also proved that
every power series solution of the KP system
should be written as
(\ref{eq4.7}), allowing certain infinite determinants.
A necessary background of the KP theory can be 
found in \cite{Mulase1994}.

We have thus proved the following theorem. 
\begin{thm}\label{thm4.2}
If $\phi_j(k)$, $j=0, \cdots, n-1$,
 satisfies that
$$
\left|\int_{-\infty} ^{\infty} k^r   \phi_j(k)  dk
\right|<+\infty
$$
for all $r\ge 0$, then the asymptotic expansion
of the matrix integral $Z_n(t,m,\phi)$ satisfies
the KP equations with respect to $T_1, T_2, \cdots,
T_{2m}$. Moreover, if we choose a value of $T_{2m}$
such that $Re(T_{2m})<0$ and fix it, then $Z_n(t,m,\phi)$
itself is an entire  holomorphic solution to the KP equations
with respect to
$(T_1, T_2, \cdots, T_{2m-1})\in {{\mathbb{C}}}^{2m-1}$.
In particular, 
$$
u(T_1, T_2, T_3, \cdots) = 
\frac{\partial^2}{\partial T_1 ^2}
\log(Z_n(t,m,\phi))
$$
is a meromorphic solution to the KP equation
$$
\frac{3}{4} u_{22} = 
\left( u_3 - \frac{1}{4} u_{111} -3u u_1
\right)_1 ,
$$
where  $u_j$ denotes the partial derivative of $u$
with respect to $T_j$.
\end{thm}

The formula we have just established
is a continuum version of the famous Hirota soliton solution
of the KP equations \cite{Sato}. The most general
soliton  solution of the KP equations due to Mikio and 
Yasuko Sato
depends on $nM+M$ parameters $c_{ij}$ and $\lambda_i$, where
$0\le i \le M-1$ and $0\le j\le n-1$. Let
$$
\eta(T, k) = \sum_{\alpha=1} ^{2m} T_\alpha  k^\alpha .
$$
Then Sato-Sato's soliton solution is given by
\begin{multline*}
\sum_{0\le i_0< \cdots < i_{n-1}\le M-1}           
\exp\left(\sum_{j=0} ^{n-1} \eta(T,\lambda_{i_j})  \right)\\
\Delta(\lambda_{i_0}, \cdots, \lambda_{i_{n-1}})
\det
\pmatrix
c_{i_0 0} &\hdots & c_{i_0 n-1}\\
\vdots && \vdots\\
c_{i_{n-1} 0} & \hdots & c_{i_{n-1} n-1}
\endpmatrix .
\end{multline*}
This coincides with our $Z_n(t,m,\phi)$
 if we take
$$
\phi_j(k) = \sum_{i=0} ^{M-1} c_{ij}  \delta(k-\lambda_i) .
$$
Therefore,  our matrix integral $Z_n(t,m,\phi)$
of (\ref{eq4.1}) with (\ref{eq4.2}) is indeed
a \emph{continuum soliton solution} of the KP equations.

So far we have dealt with the matrix integrals
with a fixed integer $m$ in this
section. As before, we can 
take the limit $m\rightarrow \infty$ of 
these integrals, which
gives formal power series 
solutions of the whole hierarchy of the 
KP equations. Note that the determinant 
expression of (\ref{eq4.6}) does not have any
explicit mention on the integer $m$. Therefore,
we have obtained the third asymptotic formula for the
matrix integral:
\begin{equation}\label{eq4.8}
\lim_{m\rightarrow \infty}
{{\mathcal{A}}}\left(Z_n(t,m,\phi)\right)
= n!\cdot \frac{c(n)}{N}
\det\left(\sum_{r = 0} ^\infty p_{r-i}(T)
\int_{-\infty} ^\infty
  k^r  \phi_j(k)  dk
\right) .
\end{equation}

\section{Transcendental solutions of the KP equations
and the Grassmannian}
\label{grass} 

There are several different ways to construct
solutions to the KP equations. 
The Krichever construction and its
generalizations  are based on the 
correspondence between certain points of the
Grassmannian of Sato \cite{Sato} and 
the algebro-geometric data consisting of an irreducible
algebraic curve (possibly singular) and a 
torsion-free sheaf on it \cite{Mulase1990}. 
These solutions deserve to be called \emph{algebraic},
because they carry geometric information of
algebraic curves. Let us call a solution to the KP equations
\emph{transcendental} if no algebraic curve 
corresponds to this solution. The natural question we
can ask is: how can we construct a transcendental 
solution? 

In this section we show that the Hermitian 
matrix integrals we have been dealing with in the
earlier sections are indeed transcendental solutions. 

The technique we show that these matrix integrals are
transcendental solutions is based on the observation 
that the points of the Grassmannian corresponding to 
these solutions satisfy a peculiar $sl(2)$ stability
condition. Since these solutions are deeply
related to the moduli theory of Riemann surfaces, the
appearance of $sl(2)$ is mysteriously
 suggestive. At present
we do not have any geometric explanation of 
the relation between the KP equations, the $sl(2)$ stability
on the Grassmannian, and the moduli theory of 
pointed Riemann surfaces.  

Let $V= \mathbb{C}((z))$ denote the field of 
formal Laurent series in one variable $z$. We fix its
\emph{polarization}
\begin{equation}
\label{eq: polarization}
\mathbb{C}((z)) = \mathbb{C}[z^{-1}] \oplus
\mathbb{C}[[z]]\cdot z .
\end{equation}
For a vector subspace $W\subset V$, there is a 
natural map
\begin{equation}
\label{eq: gamma W}
\gamma_W: W\hookrightarrow V \longrightarrow
V/\mathbb{C}[[z]] z \cong \mathbb{C}[z^{-1}] .
\end{equation}
The infinite-dimensional Grassmannian is defined by
\begin{equation}
\label{eq: infinite grassmannian}
Gr = \{W\subset V\;|\;
\gamma_W: W \longrightarrow
 \mathbb{C}[z^{-1}] {\text{ is Fredholm of index }} 0\}.
\end{equation}
The \emph{big-cell} of the 
Grassmannian is the
subset of $Gr$ consisting of vector subspaces
$W\subset V$ such that $\gamma_W$ 
of (\ref{eq: gamma W}) is an isomorphism.

Let $W$ be a point of the big-cell of the 
Grassmannian. We can choose a basis 
$$
\langle w_0, w_1, w_2, \cdots \rangle
$$ 
for $W$ such that
\begin{equation}
\label{eq: basis element}
w_j = z^{-j} + \sum_{i=1} ^{\infty}
c_{ij} z^i, \qquad j=0,1,2,\cdots .
\end{equation}
The \emph{Bosonization} is a map
\begin{equation}
\label{eq: bosonization}
Gr \longrightarrow 
\mathbb{P}(\mathbb{C}[[T_1,T_2,T_3,\cdots]])
\end{equation}
that assigns a $\tau$-function $\tau_W$ to each point
$W$ of the Grassmannian. For a point $W$ of the big-cell
with a basis (\ref{eq: basis element}), 
the Bosonization has an infinite determinant expression
\begin{equation}
\label{eq: tau of W}
\tau_W = \det\left( p_{i-j}(T) +  \sum_{\mu = 1} ^{\infty}
p_{\mu + i}(T) c_{\mu j} \right).
\end{equation}
The infinite determinant gives a well-defined
element of $\mathbb{C}[[T_1,T_2,T_3,\cdots]]$
in the same manner as we have explained in the earlier 
sections. Sato's formula (\ref{eq4.7}) gives another
expression of the Bosonization map. 
For more detail, we refer to 
\cite{Mulase1994} and \cite{Mulase1994b}.

The \emph{commutative stabilizer} of 
$W\in Gr$ is defined by
\begin{equation}
\label{eq: stabilizer}
A_W = \{ a\in \mathbb{C}((z))\;|\; a\cdot W
\subset W\}.
\end{equation} 
The key idea that connects the KP equations and
algebraic curves is that the commutative 
stabilizer is the coordinate ring of an algebraic 
curve. If the greatest common divisor of the 
pole order of elements in $A_W$ is $1$, then the
Bosonization $\tau_W$ of $W$ is essentially the Riemann
theta function associated with the algebraic 
curve $C$ whose coordinate ring is $A_W$ 
\cite{Mulase1984}, \cite{Mulase1994}. 

\begin{Def}
\label{def: transcendental}
A solution of the KP equations $\tau_W$ is said to be
\emph{transcendental} if 
\begin{equation}
\label{eq: transcendental}
A_W = \mathbb{C}.
\end{equation}
\end{Def}

\begin{rem}
It is known that if $A_W \ne \mathbb{C}$, then
the Bosonization $\tau_W$ of $W$ is a solution
to the KP equation corresponding to
a vector bundle 
$\mathcal{F}$ 
on an  algebraic curve $C$ such that
\begin{equation}
\label{eq: cohomology}
H^0(C,\mathcal{F}) = H^1(C, \mathcal{F}) = 0
\end{equation}
\cite{Mulase1990}. Conversely, there
is a solution corresponding to an arbitrary 
torsion-free sheaf $\mathcal{F}$ defined on an
arbitrary (possibly singular) algebraic curve
$C$ satisfying (\ref{eq: cohomology}).
 None of these solutions are
transcendental. 
\end{rem}

The Hermitian matrix integral we have discussed in
Section~\ref{matrix} gives a transcendental 
solution to the KP equations. 

\begin{thm}
\label{thm: transcendental solution}
Choose  arbitrary positive integers $k$ and  $n$, and
let 
$$
a = (a_1, a_2, \cdots, a_{2k}) \in 
\mathbb{C}^{2k}
$$
 be a complex vector such that
$Re(a_{2k}) < 0$. 
Define a formal Laurent series
\begin{equation}
\label{eq: wj}
w_j = \sum_{r=0} ^{\infty}
\left(
\int_{-\infty} ^{\infty}
\lambda^{r+j} \exp
\left( \sum_{\mu = 1} ^{2k} a_{\mu} \lambda^{\mu} 
\right)
d\lambda
\right)
z^{r+1-n} \in \mathbb{C}((z))
\end{equation}
for $j=0,1,2,\cdots,n-1$, and let
\begin{equation}
\label{eq: W(a)}
W(a) = \langle
w_0, w_1, \cdots, w_{n-1}, z^{-n}, z^{-n-1}, 
\cdots\rangle 
\in Gr
\end{equation}
be a point of the Grassmannian
spanned by $w_0, w_1, \cdots, w_{n-1}$, 
and $z^{-n}, z^{-n-1}, 
\cdots$. 
Then the $\tau$-function corresponding to 
$W(a)$ is given by the asymptotic
expansion of a Hermitian matrix integral:
\begin{equation}
\label{eq: matrix inegral with a}
\tau_{W(a)} = 
\lim_{m\rightarrow \infty}
\mathcal{A}\left(
\int_{\mathcal{H}_n}
\exp\left(
\sum_{j=1} ^{2m} T_j \trace(X^j)
\right)
\exp\left(
\sum_{\mu =1} ^{2k} a_{\mu} \trace(X^{\mu})
\right)
dX \right),
\end{equation}
where  we take $Re(T_{2m}) <0$ first 
and then let $m\rightarrow \infty$ to determine a
well-defined formal power series
in $\mathbb{C}[[T_1,T_2,T_3,\cdots]]$.
Define a linear differential operator
\begin{equation}
\label{eq: diff op}
L_i(a) =
z^{1-i}
\frac{d}{dz} + \frac{(3n-1) + i(n-1)}{2} z^{-i}
+ \sum_{\mu=1} ^{2k} \mu a_{\mu} z^{-i-\mu}
\end{equation}
for $i = -1, 0, 1$. These differential operators
satisfy the $sl(2, \mathbb{C})$ relation
$$
[L_i(a), L_j(a)] = (i-j) L_{i+j}(a).
$$
The point $W(a)$ of the Grassmannian satisfies the
non-commutative stability condition
\begin{equation}
\label{eq: sl2 stability}
L_i(a) \cdot W(a) \subset W(a), \qquad i=-1, 0, 1.
\end{equation}
Moreover, $\tau_{W(a)}$ is a transcendental solution
of the KP equations.
\end{thm}

\begin{proof}
The function 
$$
\exp\left(
\sum_{\mu =1} ^{2k} a_{\mu} \trace(X^{\mu})
\right)
$$
is a special case of the function 
$\phi(X)$ defined in (\ref{eq4.2}). Thus the results of
the previous section proves that 
$\tau_{W(a)}$ is a $\tau$-function of the KP
equations corresponding to the point of the
Grassmannian $W(a)$. 

Let us first prove that the $sl(2)$ stability condition
(\ref{eq: sl2 stability}) implies that the commutative
stabilizer is trivial:
$$
A_{W(a)} = \mathbb{C}.
$$
Suppose $f(z) \in A_{W(a)}\subset \mathbb{C}((z))$,
and let $\ord(f) = \nu > 0$, where we define the
\emph{pole order} by
$$
\ord(z^{-\nu}) = \nu.
$$
 Since $L_{-1}(a)$ and 
$f$ stabilize $W(a)$, 
$$
[L_{-1}(a),f] = z^2 \frac{df}{dz} \in A_{W(a)}
$$ 
also stabilizes $W(a)$. Note that 
$$
\ord([L_{-1}(a),f]) = \nu -1.
$$
Thus we can immediately conclude that 
$$
A_{W(a)} = \mathbb{C}[z^{-1}].
$$
But then
\begin{equation}
\label{eq: new stab}
L_{-1}(a) - \sum_{\mu = 1} ^{2k}
\mu a_{\mu} z^{1-\mu}
= z^2\frac{d}{dz} + 
\frac{(3n-1) -(n-1)}{2} z
\end{equation}
stabilizes $W(a)$. Since the new stabilizer
 (\ref{eq: new stab}) decreases the order of elements
of $W(a)$ exactly by $1$, $W(a)$ must have an
element of arbitrary negative order. But this 
contradicts to the Fredholm condition of $W(a)$. 
This means $A_{W(a)} = \mathbb{C}$, hence
$\tau_{W(a)}$ is a transcendental solution. 

Now all we need is to show (\ref{eq: sl2 stability}),
which can be verified by a straightforward computation.
First, we note a simple formula
\begin{equation}
\label{eq: int by parts}
\begin{split}
0&= \int_{-\infty} ^{\infty}
\frac{d}{d\lambda}\left(
\lambda^{\alpha} \exp\left(
\sum_{\mu=1} ^{2k} a_{\mu} \lambda^{\mu}\right)
d\lambda\right)\\
&= \int_{-\infty} ^{\infty}
\alpha
\lambda^{\alpha -1} \exp\left(
\sum_{\mu=1} ^{2k} a_{\mu} \lambda^{\mu}\right)
d\lambda\\
&\quad +\int_{-\infty} ^{\infty}
\sum_{\mu=1} ^{2k} \mu a_{\mu} \lambda^{\alpha + \mu-1}
 \exp\left(
\sum_{\mu=1} ^{2k} a_{\mu} \lambda^{\mu}\right)
d\lambda.\\
\end{split}
\end{equation}
Let us compute the effect of the differential operators
(\ref{eq: diff op}) on the basis elements of $W(a)$. 
First, we have
\allowdisplaybreaks
\begin{multline*}
L_{-1}(a) w_j
= \left( z^2 \frac{d}{dz} + nz + \sum_{\mu=1} ^ {2k}
\mu a_{\mu} z^{1-\mu}\right)
\sum_{r=0} ^{\infty}z^{r+1-n}
 \int_{-\infty} ^{\infty}
\lambda^{r+ j} e^{\sum_{\mu=1} ^ {2k}
 a_{\mu}\lambda^{\mu}} d\lambda\\
=\sum_{r=0} ^{\infty} (r+1) z^{r+2-n}
 \int_{-\infty} ^{\infty}
\lambda^{r+ j} e^{\sum_{\mu=1} ^ {2k}
 a_{\mu}\lambda^{\mu}} d\lambda\\
\qquad + \sum_{r=0} ^{\infty}
\sum_{\mu=1} ^{2k} z^{r+2-n-\mu}
 \int_{-\infty} ^{\infty}
\lambda^{r+ j} \mu a_{\mu}
e^{\sum_{\mu=1} ^ {2k}
 a_{\mu}\lambda^{\mu}} d\lambda\\
=\sum_{r=0} ^{\infty} r z^{r+1-n}
 \int_{-\infty} ^{\infty}
\lambda^{r+ j-1} 
e^{\sum_{\mu=1} ^ {2k}
 a_{\mu}\lambda^{\mu}} d\lambda\\
\qquad + \sum_{\mu=1} ^{2k} 
\sum_{r=0} ^{\mu-2}
z^{r+2-n-\mu}
 \int_{-\infty} ^{\infty}
\lambda^{r+ j} \mu a_{\mu}
e^{\sum_{\mu=1} ^ {2k}
 a_{\mu}\lambda^{\mu}} d\lambda\\
\qquad + \sum_{\mu=1} ^{2k} 
\sum_{r=\mu-1} ^{\infty} z^{r+2-n-\mu}
 \int_{-\infty} ^{\infty}
\lambda^{r+ j} \mu a_{\mu}
e^{\sum_{\mu=1} ^ {2k}
 a_{\mu}\lambda^{\mu}} d\lambda\\
=\sum_{r=0} ^{\infty} r z^{r+1-n}
 \int_{-\infty} ^{\infty}
\lambda^{r+ j-1} 
e^{\sum_{\mu=1} ^ {2k}
 a_{\mu}\lambda^{\mu}} d\lambda\\
\qquad + \sum_{\mu=1} ^{2k} 
\sum_{r=0} ^{\mu-2}
z^{r+2-n-\mu}
 \int_{-\infty} ^{\infty}
\lambda^{r+ j} \mu a_{\mu}
e^{\sum_{\mu=1} ^ {2k}
 a_{\mu}\lambda^{\mu}} d\lambda\\
\qquad + \sum_{\mu=1} ^{2k} 
\sum_{r=0} ^{\infty} z^{r+1-n}
 \int_{-\infty} ^{\infty}
\lambda^{r+ j+\mu -1} \mu a_{\mu}
e^{\sum_{\mu=1} ^ {2k}
 a_{\mu}\lambda^{\mu}} d\lambda\\
=\sum_{r=0} ^{\infty} r z^{r+1-n}
 \int_{-\infty} ^{\infty}
\lambda^{r+ j-1} 
e^{\sum_{\mu=1} ^ {2k}
 a_{\mu}\lambda^{\mu}} d\lambda\\
\qquad + \sum_{\mu=1} ^{2k} 
\sum_{r=0} ^{\mu-2}
z^{r+2-n-\mu}
 \int_{-\infty} ^{\infty}
\lambda^{r+ j} \mu a_{\mu}
e^{\sum_{\mu=1} ^ {2k}
 a_{\mu}\lambda^{\mu}} d\lambda\\
\qquad - \sum_{r=0} ^{\infty} z^{r+1-n}
 \int_{-\infty} ^{\infty} (r+j)
\lambda^{r+ j -1} 
e^{\sum_{\mu=1} ^ {2k}
 a_{\mu}\lambda^{\mu}} d\lambda\\
= -jw_{j-1} +  \sum_{\mu=1} ^{2k} 
\sum_{r=0} ^{\mu-2}
z^{r+2-n-\mu}
 \int_{-\infty} ^{\infty}
\lambda^{r+ j} \mu a_{\mu}
e^{\sum_{\mu=1} ^ {2k}
 a_{\mu}\lambda^{\mu}} d\lambda\\
\in W(a)\\
\end{multline*}
for all $j=0, 1, 2, \cdots, n-1$. Note that $w_{-1}$
does not appear in the above computation because 
of the combination $jw_{j-1}$. For the basis elements
$z^{-n}, z^{-n-1}, \cdots,$ we have
\begin{equation*}
\begin{split}
L_{-1}(a) z^{-n-i}
&= \left( z^2 \frac{d}{dz} + nz + \sum_{\mu=1} ^ {2k}
\mu a_{\mu} z^{1-\mu}\right) z^{-n-i}\\
&= (-i) z^{-n-i+1} + \sum_{\mu=1} ^{2k}
\mu a_{\mu} z^{1-\mu-n-i}
\in W(a)\\
\end{split}
\end{equation*}
for all $i\ge 0$. We note that the term $z^{-n+1}$
does not appear in this computation. Thus we conclude
$$
L_{-1}(a)\cdot W(a) \subset W(a).
$$
For $j=0$, we have
\begin{multline*}
L_0 (a) w_j
= \left( z \frac{d}{dz} + \frac{3n-1}{2} + 
\sum_{\mu=1} ^ {2k}
\mu a_{\mu} z^{-\mu}\right)
\sum_{r=0} ^{\infty}z^{r+1-n}
 \int_{-\infty} ^{\infty}
\lambda^{r+ j} e^{\sum_{\mu=1} ^ {2k}
 a_{\mu}\lambda^{\mu}} d\lambda\\
=\sum_{r=0} ^{\infty} (r+\frac{n+1}{2}) z^{r+1-n}
 \int_{-\infty} ^{\infty}
\lambda^{r+ j} e^{\sum_{\mu=1} ^ {2k}
 a_{\mu}\lambda^{\mu}} d\lambda\\
\qquad + \sum_{r=0} ^{\infty}
\sum_{\mu=1} ^{2k} z^{r+1-n-\mu}
 \int_{-\infty} ^{\infty}
\lambda^{r+ j} \mu a_{\mu}
e^{\sum_{\mu=1} ^ {2k}
 a_{\mu}\lambda^{\mu}} d\lambda\\
=\frac{n+1}{2} w_j + \sum_{r=0} ^{\infty} r z^{r+1-n}
 \int_{-\infty} ^{\infty}
\lambda^{r+ j} 
e^{\sum_{\mu=1} ^ {2k}
 a_{\mu}\lambda^{\mu}} d\lambda\\
\qquad + \sum_{\mu=1} ^{2k} 
\sum_{r=0} ^{\mu-1}
z^{r+1-n-\mu}
 \int_{-\infty} ^{\infty}
\lambda^{r+ j} \mu a_{\mu}
e^{\sum_{\mu=1} ^ {2k}
 a_{\mu}\lambda^{\mu}} d\lambda\\
\qquad + \sum_{\mu=1} ^{2k} 
\sum_{r=\mu} ^{\infty} z^{r+1-n-\mu}
 \int_{-\infty} ^{\infty}
\lambda^{r+ j} \mu a_{\mu}
e^{\sum_{\mu=1} ^ {2k}
 a_{\mu}\lambda^{\mu}} d\lambda\\
=\frac{n+1}{2} w_j +\sum_{r=0} ^{\infty} r z^{r+1-n}
 \int_{-\infty} ^{\infty}
\lambda^{r+ j} 
e^{\sum_{\mu=1} ^ {2k}
 a_{\mu}\lambda^{\mu}} d\lambda\\
\qquad + \sum_{\mu=1} ^{2k} 
\sum_{r=0} ^{\mu-1}
z^{r+1-n-\mu}
 \int_{-\infty} ^{\infty}
\lambda^{r+ j} \mu a_{\mu}
e^{\sum_{\mu=1} ^ {2k}
 a_{\mu}\lambda^{\mu}} d\lambda\\
\qquad + \sum_{\mu=1} ^{2k} 
\sum_{r=0} ^{\infty} z^{r+1-n}
 \int_{-\infty} ^{\infty}
\lambda^{r+ j+\mu } \mu a_{\mu}
e^{\sum_{\mu=1} ^ {2k}
 a_{\mu}\lambda^{\mu}} d\lambda\\
=\frac{n+1}{2} w_j +\sum_{r=0} ^{\infty} r z^{r+1-n}
 \int_{-\infty} ^{\infty}
\lambda^{r+ j} 
e^{\sum_{\mu=1} ^ {2k}
 a_{\mu}\lambda^{\mu}} d\lambda\\
\qquad + \sum_{\mu=1} ^{2k} 
\sum_{r=0} ^{\mu-1}
z^{r+1-n-\mu}
 \int_{-\infty} ^{\infty}
\lambda^{r+ j} \mu a_{\mu}
e^{\sum_{\mu=1} ^ {2k}
 a_{\mu}\lambda^{\mu}} d\lambda\\
\qquad - \sum_{r=0} ^{\infty} z^{r+1-n}
 \int_{-\infty} ^{\infty} (r+j+1)
\lambda^{r+ j } 
e^{\sum_{\mu=1} ^ {2k}
 a_{\mu}\lambda^{\mu}} d\lambda\\
= \left(\frac{n+1}{2}-j-1\right) w_j 
+\sum_{\mu=1} ^{2k} 
\sum_{r=0} ^{\mu-1}
z^{r+1-n-\mu}
 \int_{-\infty} ^{\infty}
\lambda^{r+ j} \mu a_{\mu}
e^{\sum_{\mu=1} ^ {2k}
 a_{\mu}\lambda^{\mu}} d\lambda\\
\in W(a)\\
\end{multline*}
for all $j=0, 1, 2, \cdots, n-1$. It is obvious that
$$
L_0(a)\cdot  z^{-n-i} \in W(a)
$$ 
for  $i\ge 0$. Finally, for $j=1$, we have
\begin{multline*}
L_1 (a) w_j\\
= \left( \frac{d}{dz} + (2n-1)z^{-1} + 
\sum_{\mu=1} ^ {2k}
\mu a_{\mu} z^{-\mu-1}\right)
\sum_{r=0} ^{\infty}z^{r+1-n}
 \int_{-\infty} ^{\infty}
\lambda^{r+ j} e^{\sum_{\mu=1} ^ {2k}
 a_{\mu}\lambda^{\mu}} d\lambda\\
=\sum_{r=0} ^{\infty} (r+n) z^{r-n}
 \int_{-\infty} ^{\infty}
\lambda^{r+ j} e^{\sum_{\mu=1} ^ {2k}
 a_{\mu}\lambda^{\mu}} d\lambda\\
\qquad + \sum_{r=0} ^{\infty}
\sum_{\mu=1} ^{2k} z^{r-n-\mu}
 \int_{-\infty} ^{\infty}
\lambda^{r+ j} \mu a_{\mu}
e^{\sum_{\mu=1} ^ {2k}
 a_{\mu}\lambda^{\mu}} d\lambda\\
= \sum_{r=-1} ^{\infty} (r+n+1) z^{r+1-n}
 \int_{-\infty} ^{\infty}
\lambda^{r+ j+1} 
e^{\sum_{\mu=1} ^ {2k}
 a_{\mu}\lambda^{\mu}} d\lambda\\
\qquad + \sum_{\mu=1} ^{2k} 
\sum_{r=0} ^{\mu}
z^{r-n-\mu}
 \int_{-\infty} ^{\infty}
\lambda^{r+ j} \mu a_{\mu}
e^{\sum_{\mu=1} ^ {2k}
 a_{\mu}\lambda^{\mu}} d\lambda\\
\qquad + \sum_{\mu=1} ^{2k} 
\sum_{r=\mu+1} ^{\infty} z^{r-n-\mu}
 \int_{-\infty} ^{\infty}
\lambda^{r+ j} \mu a_{\mu}
e^{\sum_{\mu=1} ^ {2k}
 a_{\mu}\lambda^{\mu}} d\lambda\\
=z^{-n}
 \int_{-\infty} ^{\infty}
\lambda^{j} 
e^{\sum_{\mu=1} ^ {2k}
 a_{\mu}\lambda^{\mu}} d\lambda
+\sum_{r=0} ^{\infty} (r+n+1) z^{r+1-n}
 \int_{-\infty} ^{\infty}
\lambda^{r+ j+1} 
e^{\sum_{\mu=1} ^ {2k}
 a_{\mu}\lambda^{\mu}} d\lambda\\
\qquad + \sum_{\mu=1} ^{2k} 
\sum_{r=0} ^{\mu}
z^{r-n-\mu}
 \int_{-\infty} ^{\infty}
\lambda^{r+ j} \mu a_{\mu}
e^{\sum_{\mu=1} ^ {2k}
 a_{\mu}\lambda^{\mu}} d\lambda\\
\qquad + \sum_{\mu=1} ^{2k} 
\sum_{r=0} ^{\infty} z^{r+1-n}
 \int_{-\infty} ^{\infty}
\lambda^{r+ j+\mu +1} \mu a_{\mu}
e^{\sum_{\mu=1} ^ {2k}
 a_{\mu}\lambda^{\mu}} d\lambda\\
=z^{-n}
 \int_{-\infty} ^{\infty}
\lambda^{j} 
e^{\sum_{\mu=1} ^ {2k}
 a_{\mu}\lambda^{\mu}} d\lambda
+\sum_{r=0} ^{\infty} (r+n+1) z^{r+1-n}
 \int_{-\infty} ^{\infty}
\lambda^{r+ j+1} 
e^{\sum_{\mu=1} ^ {2k}
 a_{\mu}\lambda^{\mu}} d\lambda\\
\qquad + \sum_{\mu=1} ^{2k} 
\sum_{r=0} ^{\mu}
z^{r-n-\mu}
 \int_{-\infty} ^{\infty}
\lambda^{r+ j} \mu a_{\mu}
e^{\sum_{\mu=1} ^ {2k}
 a_{\mu}\lambda^{\mu}} d\lambda\\
\qquad - \sum_{r=0} ^{\infty} z^{r+1-n}
 \int_{-\infty} ^{\infty} (r+j+2)
\lambda^{r+ j+1 } 
e^{\sum_{\mu=1} ^ {2k}
 a_{\mu}\lambda^{\mu}} d\lambda\\
= z^{-n}
 \int_{-\infty} ^{\infty}
\lambda^{j} 
e^{\sum_{\mu=1} ^ {2k}
 a_{\mu}\lambda^{\mu}} d\lambda
+(n-j-1) w_{j+1}\\
\qquad + \sum_{\mu=1} ^{2k} 
\sum_{r=0} ^{\mu}
z^{r-n-\mu}
 \int_{-\infty} ^{\infty}
\lambda^{r+ j} \mu a_{\mu}
e^{\sum_{\mu=1} ^ {2k}
 a_{\mu}\lambda^{\mu}} d\lambda\\
\in W(a)\\
\end{multline*}
for all $j=0, 1, 2, \cdots, n-1$.
Note that the term $w_n$ does not appear in the
computation. It is again obvious that 
$$
L_1(a)\cdot z^{-n-i}\in W(a)
$$
for $i\ge 0$. This completes the proof of the
$sl(2)$ stability of $W(a)$,
 and hence we have established the
theorem.
\end{proof}

The action 
of these $sl(2)$ generators on $W(a)$
is very subtle, and it does not
seem to allow any generalization. For example,
the above proof does not apply for the Virasoro
generators $L_i(a)$ other than $i=-1, 0, 1$, although the
operators $L_i(a)$ are defined for all $i\in \mathbb{Z}$
and they satisfy the Witt algebra relation
$$
[L_i(a), L_j(a)] = (i-j)L_{i+j}(a)
$$
for $i,j\in \mathbb{Z}$.

\bibliography{hermit}
\bibliographystyle{plain}

\end{document}